\documentclass[onecolumn,amsmath,amssymb,aps,pra,floatfix,superscriptaddress]{revtex4-1}
\usepackage[utf8]{inputenc}
\usepackage{natbib}
\usepackage{graphicx}
\usepackage{float}\usepackage[tight,footnotesize]{subfigure}
\usepackage{color}
\usepackage{mathtools}
\usepackage{mathptmx}
\usepackage[scaled=.95]{helvet}
\usepackage{courier}

\begin{document}
\title{Jamming transitions in force-based models for pedestrian dynamics}
\author{Mohcine Chraibi}
\affiliation{J\"ulich Supercomputing Centre, Forschungszentrum J\"ulich, 
  52425 J\"ulich, Germany}
\email{m.chraibi@fz-juelich.de}
\author{Takahiro Ezaki}

\affiliation{ 
  Department of Aeronautics and Astronautics, Graduate School of Engineering, The University of Tokyo, 7-3-1, Hongo, Bunkyo-ku, Tokyo 113-8656, Japan
}
\author{Antoine Tordeux}
\affiliation{J\"ulich Supercomputing Centre, Forschungszentrum J\"ulich,  
  52425 J\"ulich, Germany}
\author{Katsuhiro Nishinari}
\affiliation{Research Center for Advanced Science and Technology, 
  The University of Tokyo, 4-6-1, Komaba, Meguro-ku, Tokyo 153-8904, Japan}
\author{Andreas Schadschneider}
\affiliation{Institute for Theoretical Physics, Universit\"at zu K\"oln, 
  50937 K\"oln, Germany}
\author{Armin Seyfried}
\affiliation{J\"ulich Supercomputing Centre, Forschungszentrum J\"ulich,  
  52425 J\"ulich, Germany}


\date{\today}


\begin{abstract}
  Force-based models describe pedestrian dynamics in analogy to
  classical mechanics by a system of second order ordinary
  differential equations.  
  By investigating the linear stability of two main classes of
    forces, parameter regions with unstable homogeneous states are
  identified.  In this unstable regime it is then checked whether
    phase transitions or stop-and-go waves occur. Results based on 
    numerical simulations show, however, that the investigated models 
    lead to unrealistic behavior in form of backwards moving 
    pedestrians and overlapping. 
    This is one reason why stop-and-go waves have
      not been observed in these models. The unrealistic behavior
    is not related to the numerical treatment of the dynamic equations
    but rather indicates an intrinsic problem of this model class.
    Identifying the underlying generic problems gives indications how
    to define models that do not show such unrealistic behavior. As an
    example we introduce a new force-based model which produces
    realistic jam dynamics without the appearance of unrealistic
    negative speeds for empirical desired walking speeds.

\end{abstract}

\maketitle
\section{Introduction}

Mathematical models based on ideas from physics can improve our
understanding of the characteristics of crowds and give useful
insights into their dynamics. From a more practical point of view
  such models have applications e.g.\ in safety analysis of large
  public events where they may help predicting critical situations,
  allowing preventive measures.

A popular class of models is of microscopic nature, describing
the dynamics of crowds by specifying properties of individuals and
defining their interactions. The most elaborated models belong either to the
subclass of rule-based models that are discrete in space (i.e. cellular
automata), or to
force-based models continuous in space, which are described by a
system of second order ordinary differential equations~\cite{Helbing2001,Schadschneider2009a,Schadschneider2010b,Ali2013}. 

Especially for applications in safety analysis,
models that are validated qualitatively and quantitatively are
required.  Quantitative validation of pedestrian dynamics consists of
measuring density, velocity and flow in simulations and comparing them
with empirical data. The relation between these quantities, also
called the fundamental diagram, is widely considered as the most
important criterion to validate simulation results
\cite{Seyfried2008,Schadschneider2009c}.  Besides this quantitative
validation often the focus is more on the reproduction of qualitative
properties, especially collective effects.  Most of the force-based
models are in fact able to describe fairly well some of those phenomena, 
e.g.\ lane formation~\cite{ZhangQ2011,Yu2005}, oscillations at
bottlenecks~\cite{ZhangQ2011,Helbing2004}, the ``faster-is-slower''
effect~\cite{Lakoba2005,Parisi2007} and clogging at bottlenecks
\cite{Helbing2004,Yu2005}, that sometimes are difficult to verify
empirically~\cite{Garcimartin2014,Parisi2015}.

An often observed collective phenomenon
that emerges in crowds, especially when the density exceeds a critical
value, is stop-and-go waves~\cite{Schadschneider2009a}.  Although some
space-continuous models
\cite{Portz2010,Seyfried2010b,Lemercier2012,Eilhardt2014} reproduce
partly this phenomenon, force-based models generally fail to describe
pedestrian dynamics in jam situations correctly.  Instead in some situations
quite often unrealistic behavior like backward motion or overtaking
(``tunneling'') is observed, especially in one-dimensional single-file
scenarios. Recently it has been shown~\cite{Chraibi2014a} that this is
not a consequence of numerical problems in the treatment of the
differential equations, but an indication of inherent problems of
force-based models, at least for certain classes of forces.

In vehicular traffic, the formation of jams and the dynamics of
traffic waves have been studied intensively~\cite{Chowdhury2000,Gazis2002,Orosz2010,Nagatani2002a}.  
Traffic jams in simulations occur as a result of phase transitions from a
stable homogeneous configuration to an unstable configuration. 
That means  it should be possible to calibrate model parameters such that
systems in  unstable regimes can be simulated.
Otherwise, a reproduction of jams is impossible and the model can be
qualified as unrealistic.  For each parameter set that leads to an
unstable homogeneous state it has to be verified by simulations
whether this instability corresponds to realistic behavior (i.e.~the
occurrence of jams) or unrealistic behavior (e.g. overlapping of
particles).  A certain amount of overlapping might be acceptable as it
could be interpreted as ``elasticity'' of the particles.  Generically,
however, the amount of overlapping is not limited in these models and
even ``tunneling'' of particles is observed.

In pedestrian dynamics, numerous force-based models have been
developed based on physical analogies, i.e. Newtonian dynamics.
Pedestrian dynamics is described as a deviation from a predefined
desired direction resulting from forces acting on each pedestrian.
These forces are not fundamental physical forces, but effective forces
that give a physical interpretation of the decisions made by
pedestrians.  Therefore the forces can not be measured directly but
only via their effects on the motion, i.e.\ the observed
accelerations.  This might be one reason why in the literature a
diversity of models has been proposed, e.g.\ based on algebraically
decaying forces, exponential forces etc. 
Although the force-based
Ansatz is elegant and to some extent helpful in describing the
dynamics of pedestrians, it has some intrinsic problems that we will
discuss in this paper. 
These problems were observed earlier and have lead to modifications
of the original models by introducing additional forces, like a physical 
force, or even restrictions on the state variables. 

K\"oster et al.~\cite{Koester2013} gave a thorough analysis of the
numerical problems that are encountered when simulating pedestrian
dynamics with force-based models.  As shown in~\cite{Koester2013,Chraibi2014}
 the problem of oscillations in the
trajectories of pedestrian (backwards movement) is an
\textit{intrinsic} problem of second order models, not (only)
 a numerical one due
 to the accuracy of the numerical solver.  In~\cite{Chraibi2014a} an analytical investigation of the social
 force model in one-dimensional space proved that oscillations can only be avoided by choosing 
 values in some defined parameter spaces.  Unfortunately, these parameter
 values are either unrealistic (if they have a physical meaning) or
 they lead to large amount of overlapping (and in extreme cases, e.g.\ 
 high densities, tunneling) of pedestrians.  This so-called
 overlapping-oscillation duality is discussed in more detail
 in~\cite{Chraibi2010a,Chraibi2011}.  These problems that often lead
 to a ``complexification'' of the original (elegant) Ansatz of
 force-based models, may explain the paradigm shift observed lately
 with the emergence of new first-order models or so called ``velocity
 models''~\cite{Berg2008,Maury2009,Venel2010,Patil2010,Portz2010,Dietrich2014,Dietrich2014a,Eilhardt2014,Kirik2014}.

In this work we introduce a classification of force-based models
according to the form of the repulsive force.  The stability
properties of each class can be investigated separately in a
  unified way. Analytical criteria that ensure reproduction of
stop-and-go waves in terms of the instability of uniform single-file
motion are derived. 
Furthermore, we
analyze the influence of specific parameters of the overall behavior
of the investigated model.  A focus is on the analytical forms of the
models, and not on eventual numerical difficulties.  Based on
numerical simulations we show that the investigated models behave
unrealistically in unstable regimes, which is manifested in negative
speeds (movement in the opposite of the desired direction) and
oscillations in position of pedestrians (leads to nonphysical
overlapping).  After identifying the origin of this unrealistic
  behavior we attempt to develop a new model that mitigates these
  problems. We observe that this model shows phase separation in its
  unstable regime, in agreement with empirical results~\cite{Portz2011}. We
conclude with a discussion of the results and analysis of their
consequences as well as a detailed discussion of the limitation of the
proposed model in special and force-based models in general.


\section{Model Definition}

Pedestrian dynamics is generically a two-dimensional problem.
In order to reduce the complexity and to capture the essentials of the
jamming dynamics, we focus here on 1D systems. 
Furthermore we assume an asymmetric nearest-neighbor interaction where
the motion of a pedestrian is only influenced by the person
immediately in front.  $N$ pedestrians are initially distributed
uniformly in a one-dimensional space with periodic boundary
conditions.  Important information can then be derived from
the reaction of the system in the uniform steady state to small perturbations.

\begin{figure}[h!]
  \centering
  \includegraphics[width=0.75\columnwidth]{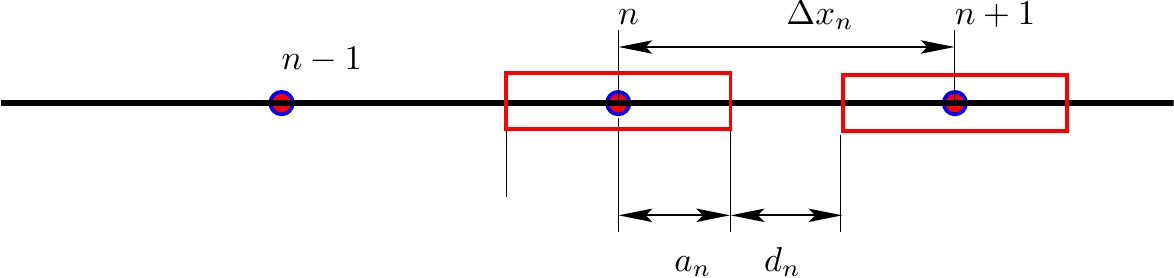}
  \caption{
    (Color online) Definition of the quantities characterizing the 
    single-file motion} of pedestrians (represented by rectangles).
  \label{fig1}
\end{figure}

For the state variables position $x_n$ and velocity $\dot
x_n=\frac{dx_n}{dt}$ of pedestrian $n$ we define the distance of the
centers $\Delta x_n$ and the relative velocity $\Delta\dot x_n$ of two
successive pedestrians, respectively, as (see Fig.~\ref{fig1})
\begin{equation}
  \Delta  x_n = x_{n+1} -  x_n,\qquad
  \Delta\dot x_n = \dot x_{n+1} -  \dot x_n\,.
\end{equation}

For convenience, we will mainly use dimensionless quantities
in the following. These are defined by the transformation
\begin{equation} 
  t\rightarrow t'=\frac{t}{\tau}\qquad \text{and}\quad 
  x_n \rightarrow x_n' = \frac{x_n}{a_0}\,,
  \label{eq-rescale}
\end{equation}
with time constant $\tau$ and the length constant $a_0$.
To simplify the notation
we denote the rescaled velocity by $\dot x_n'=\mbox dx_n'/\mbox dt'$.

In general, pedestrians are modeled as simple geometric objects of
constant size, e.g.\ a circle or ellipse.  In one-dimensional space
the size of pedestrians is characterized by $a_n$ (Fig.~\ref{fig1}),
i.e.\ their length is $2a_n$. However, it is well-known that the space
requirement of a pedestrian depends on its velocity and is defined in
a general way as a linear function of the velocity~\cite{Weidmann1993}
\begin{equation}
  a_n = a_0+a_v \dot x_n\,.
  \label{eq:an}
\end{equation}
In the following, the parameter $a_0$, characterizing the space
requirement of a standing person, will be used as length
scale for the dimensionless quantities (\ref{eq-rescale}). 
Note that the parameter $a_v\ge0$ has the dimension of time.
The dimensionless spacing $a_n' =a_n/a_0$ is written as
\begin{equation}
  a_n' = 1 + \tilde a_v \dot x_n'\,, \qquad \text{with}\quad
  \tilde a_v=\frac{a_v}{\tau}\,.
  \label{eq:an'}
\end{equation}
The effective distance (distance gap) $d_n$ of two consecutive
pedestrians becomes in dimensionless form
\begin{equation}                        
  d_n^\prime = \frac{d_n}{a_0} = \Delta x_n' - a_n'-a_{n+1}'
  =\Delta x_n'  - \tilde a_v\left(\dot x_n' +\dot x_{n+1}'\right) - 2.  
  \label{eq:effDistn}                 
\end{equation}

The dynamical equation of force-based models is usually defined as the
superposition of a repulsive force $f$ and a driving term
$g$~\cite{Chraibi2012a}.  The driving term is of central
  importance and the standard form used is
\begin{equation}
  g(\dot x_n)=\frac{v_0-\dot x_n}{\tau} \,.
  \label{eq:frv}
\end{equation}
Typical values for the parameters are $\tau=0.5$~s for the relaxation
time and $v_0=1.2$~m/s for the desired speed. Note that $\tau$ is the
same time scale used in Eq.~(\ref{eq-rescale}).
This definition gives rise to exponential acceleration to $v_0$
in free-flow movement. 
The equation of motion for pedestrian $n$ has the generic form
\begin{equation}
  \ddot x_n = f\Big(\dot x_n, \Delta \dot x_n, \Delta
  x_n\Big) + g(\dot x_n )\cdot
  \label{eq1}
\end{equation}

In this work we limit ourselves to models that incorporate
(\ref{eq:frv}) as driving term and investigate the stability of
several force-based models, defined through different 
functions $f(\cdot)$ corresponding to repulsive forces that 
either decay algebraically or exponentially with distance. 
We consider uni-dimensional dynamics and totally asymmetric 
interaction with the predecessor and assume that the repulsive 
forces are negative.
We determine their instability regions where the
investigated model may be able to reproduce stop-and-go waves. 
Technical details of the stability analysis, which is a standard
tool that can lead to cumbersome calculations, are deferred to
appendix which provides all relevant results.



\section{Models with algebraically decaying forces}

\begin{figure}[H]
  \centering
  \includegraphics[width=0.35\columnwidth]{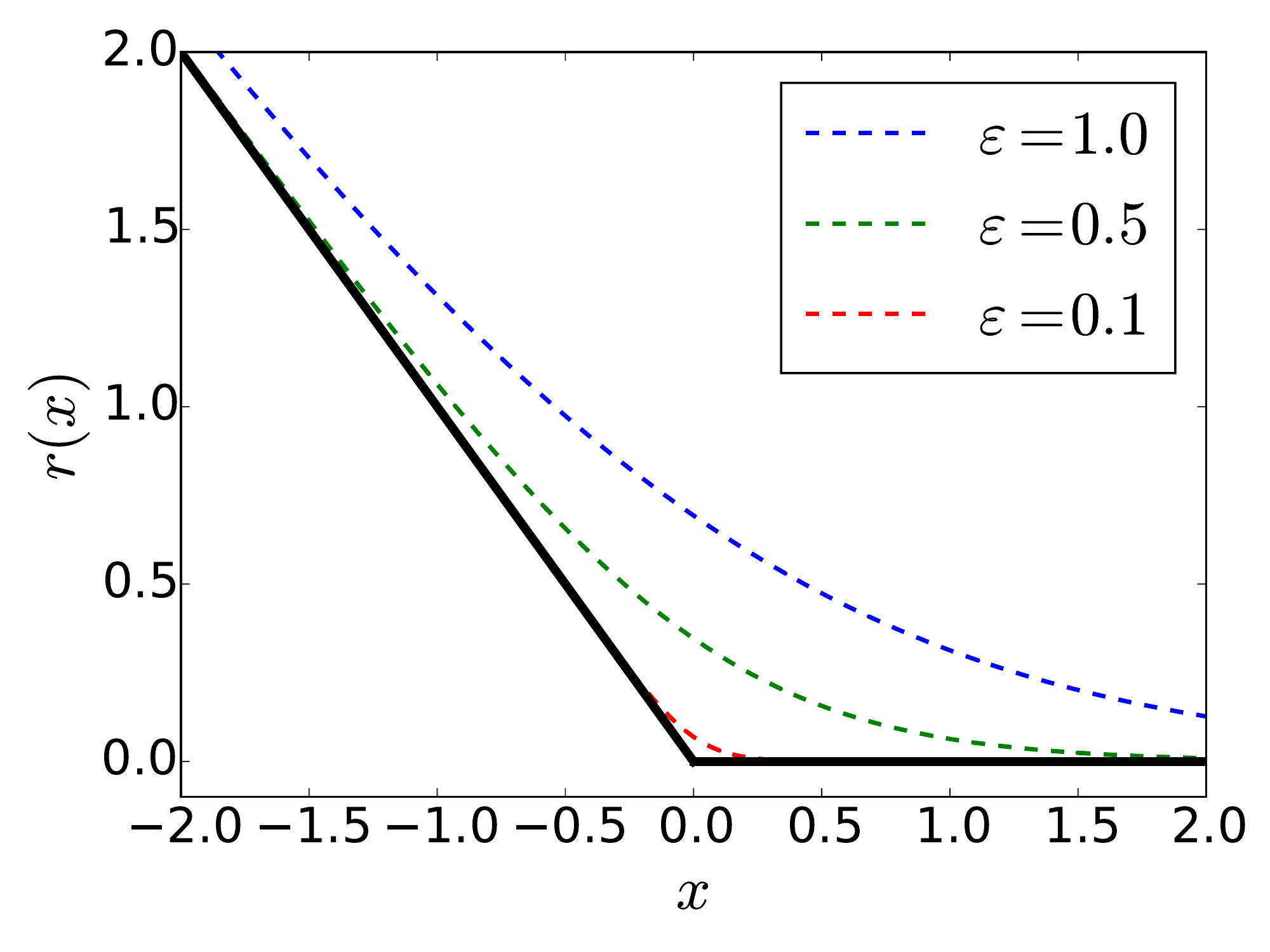}
\vspace{-0.4cm}
  \caption{(Color online) Approximations $r_\varepsilon(\cdot)$ of the ramp
    function $r(\cdot)$ (thick line). In the simulations we use $\varepsilon=0.1$.}
  \label{fig:theta}
\end{figure}

In this section we consider force-based models with an
algebraically decaying repulsive term, i.e., 
\begin{equation}
  f\Big(\dot x_n, \Delta \dot x_n,
  \Delta x_n\Big) \propto 1/(d_{n}) ^q.
  \label{eqmain}
\end{equation}
More specifically we consider the following dimensionless 
equation of motion:
\begin{equation}
  \ddot x_n' = - \frac{\Big(\mu + \delta\cdot r_\varepsilon(\Delta \dot
    x_n')\Big)^2}{{d_{n}^\prime}^q} + v_0'-
  \dot x_n',
  \label{eq:invd}
\end{equation}
with a dimensionless parameter $\mu\ge 0$ to adjust the strength of
the force, the dimensionless desired speed
$v_0'=\frac{v_0\tau}{a_0}>0$ and constants $\delta\ge 0$ and $q>0$.
In two-dimensional space the case $q<1$ corresponds to a long-ranged
repulsive force, whereas the force is short-ranged for $q>1$. 
Note that the definition of the model implies that each pedestrian only interacts with its predecessor.
Eq.~(\ref{eq:invd}) can be interpreted as extension of the generalized centrifugal force
model~\cite{Chraibi2010a} which corresponds to the
special case $\delta=1$.  The differentiable function
\begin{equation}
  r_\varepsilon(x)=\varepsilon \log(1+e^{-x/\varepsilon})
  \qquad
  (0 < \varepsilon \ll 1),
  \label{eq:aTheta}
\end{equation}
is an approximation of the non-differentiable ramp function
\begin{equation}
  r(x)=
  \begin{cases} 0, & x \ge 0, \\ -x, & \rm{else}, \end{cases}
\end{equation}
as $\varepsilon\rightarrow0$ (see Fig.~\ref{fig:theta}).  
This function suppresses the repulsive effect of a 
predecessor moving faster than the follower.
(We will set $\varepsilon=0.1$ in the simulations). 

\subsection{Model Classification}

The model class defined by Eq.~(\ref{eq:invd}) depends on
four (dimensionless) parameters $\mu$, $\delta$, $q$, and $\tilde a_v$
[which enters via (\ref{eq:effDistn})] 
and includes several models studied previously.
In the following each model will by specified by
the quadruple
$\mathcal{Q}=\langle\mu, \delta, q, \tilde a_v\rangle$.  As we will
see later the parameters $\delta$ and $\tilde a_v$ are most critical
for the dynamics described by Eq.~(\ref{eq:invd}). The parameter
$\delta$ controls the influence of the relative velocity, whereas
$\tilde a_v$ determines the velocity-dependence of the effective size
of the pedestrians.  Although in principle $\delta$ can be any real
number, in most known models it takes only discrete values in $\{0,
1\}$.

In the Centrifugal Force Model (CFM)~\cite{Yu2005} the size of
the pedestrians is independent of their speed. In addition, the CFM
considers the effects of the relative velocity $\Delta\dot x_n$,
such that slow pedestrians are not effected by faster ones. Hence, we
can define the CFM as $\mathcal{Q}=\langle 0,1,1,0\rangle$.
In contrast to the CFM, the Generalized Centrifugal Force Model (GCFM)
\cite{Chraibi2010a} includes both components - the relative velocity
and the velocity-dependence of the volume exclusion~\footnote{In GCFM
  pedestrians are modeled by ellipses with two velocity-dependent
  semi-axes.}. Additionally, to avoid overlapping of pedestrians that
results from repulsive forces among pedestrians that are too
small, moving \textit{nearly} in lockstep, a non-negative constant
$\mu$ is added to the relative velocity. Thus, the GCFM
corresponds to the case $\mathcal{Q}=\langle
\mu, 1, 1,\tilde a_v\rangle$.

Another model that represents pedestrians with constant circles and
thus has $\tilde a_v=0$ was introduced in Ref.~\cite{Helbing2000a} to
which we will refer to as HFV (Helbing, Farkas, Vicsek). Different
to the CFM and GCFM, in HFV the effects of the relative velocity are
ignored so that the HFV can be characterized by $\mathcal{Q}=\langle
\mu , 0, 2,0\rangle$.  In Ref.~\cite{Seyfried2006} an enhancement of
the HFV was introduced by Seyfried et al (SEY) consisting on a
velocity-dependent space requirement, i.e.  $\mathcal{Q}=\langle \mu
\ne 0, 0, 2, \tilde a_v\ne 0\rangle$.  Furthermore, in
Refs.~\cite{Guo2010,Guo2012} Guo et al investigated a slightly
different model (GUO) with the focus on navigation in two-dimensional
space. The GUO model can be classified as $\mathcal{Q}=\langle \mu ,
0, 1, 0\rangle$.  Similar models introducing new features have been
proposed in Refs.  \cite{Lohner2010} and~\cite{Shiwakoti2011} with a
constant added to the denominator of $f(\cdot)$. They correspond to
the case $\mathcal{Q}=\langle \mu , 0, 2, 0\rangle$.

In Tab.~\ref{tab:model} a brief summary of the aforementioned models is given.
\begin{table}[H]
  \begin{center}
    \begin{tabular}{ | c || c | }
      \hline 
      Model & $\mathcal{Q}=\langle \mu, \delta, q,\tilde a_v\rangle$\\ \hline
      \hline
      CFM   & $\langle
              0,1,1,0\rangle$\\ \hline
      GCFM   &  $\langle \mu , 1, 1,\tilde a_v\rangle$\\ \hline
      HFV &   $\langle \mu, 0, 2, 0\rangle$\\ \hline
      SEY &  $\langle \mu, 0, 2, \tilde a_v\rangle$\\ \hline
      GUO  & $\langle \mu, 0, 1, 0\rangle$ \\ \hline
      \hline
    \end{tabular}
  \end{center}
  \label{tab:model}
\vspace{-0.4cm}
\caption{ $\mathcal{Q}$-values of the investigated  
models with algebraically decaying forces.}
\end{table}

Some force-based model rely on additional algorithmic solutions like
collision detection techniques~\cite{Yu2005} or a time-to-collision
constant~\cite{Karamouzas2014} that allows to manage collisions
in simulations.  Other models rely on optimization algorithms to
define the desired direction of pedestrians~\cite{Moussaid2011}
depending on the situation of every pedestrian in the simulation.
While these additional components may prove to be useful for numerical
simulations, they have the downside of adding more complexity to the
model while stretching the concept of force-based modeling beyond
  the original idea. In some models, e.g.~\cite{Karamouzas2014},
these components are strongly correlated with the forces, which
complicates the analytical investigation of the ``pure'' force model.
Therefore, in this paper the analytical investigation is limited
solely to the force-based models that can be formulated without any
additional algorithmic components.

\subsection{Linear Stability}
\label{sub-stab1}

We study the linear stability of the system (\ref{eq:invd}) for a
given set $\mathcal{Q}$ of parameters.  The positions of the
pedestrians in the homogeneous steady state are given by
\begin{equation}
  y_n=\frac1{a_0}\left(\frac{n}{\rho} + vt\right)\,, 
\end{equation}
so that
$y_{n+1}-y_n=\frac1{a_0\rho}=\Delta y$, $\dot y_n=v\tau/a_0=v'$
and $\ddot y_n=0$
for all $n$, where derivatives are taken with respect to $t'$.
Now we consider small (dimensionless) perturbations $\epsilon_n$ 
of the steady state positions,
\begin{equation}
  x_n' = y_n + \epsilon_n\cdot
  \label{eq-pertub}
\end{equation}
For perturbations of the form 
\begin{equation}
  \epsilon_n(t)=\alpha_ne^{z t},
\end{equation}
with $\alpha_n,z\in\mathbb{C}$ we then find (expanding to first order)
\begin{equation}
  z^2 = \delta\gamma\frac{e^{\boldsymbol{i}k}-1}{{d^\prime}^q}z-
  \phi \tilde a_v(e^{\boldsymbol{i}k}+1)z
  +\phi(e^{\boldsymbol{i}k}-1)-z ,
  \label{eq:stabN}
\end{equation}
with $\gamma=\mu + \delta\varepsilon\log(2)$, $\phi=
\frac{q\gamma^2}{{d^\prime}^{q+1}}$ and $k=2\pi l/N$ with
$l=0,\ldots,N-1$.  Details of the derivation can be found in
the appendix, Sec.~\ref{appA}.

For $k\approx 0 $ we can expand $z$ as a polynomial in $k$:
\begin{equation}
  z=z^{(0)}k + z^{(1)}k^2 + \cdots
\end{equation}
Up to second order we then find the stability condition
(see appendix Sec.~\ref{appA2})
\begin{equation}
  \gamma>0,\qquad \Phi \coloneqq  \phi\omega 
- \frac{\delta\gamma}{{d^\prime}^q} - \frac{1}{2}<0\cdot
  \label{condition1}
\end{equation}
Here
$d'=\Delta y-2\tilde a_vv-2$,
with $\tilde a_v=a_v/\tau$ and $\omega=1/(2\tilde a_v\phi+1)$. 

\begin{figure}[H]
  \begin{center}
    \subfigure{}
    \includegraphics[width=0.32\columnwidth]{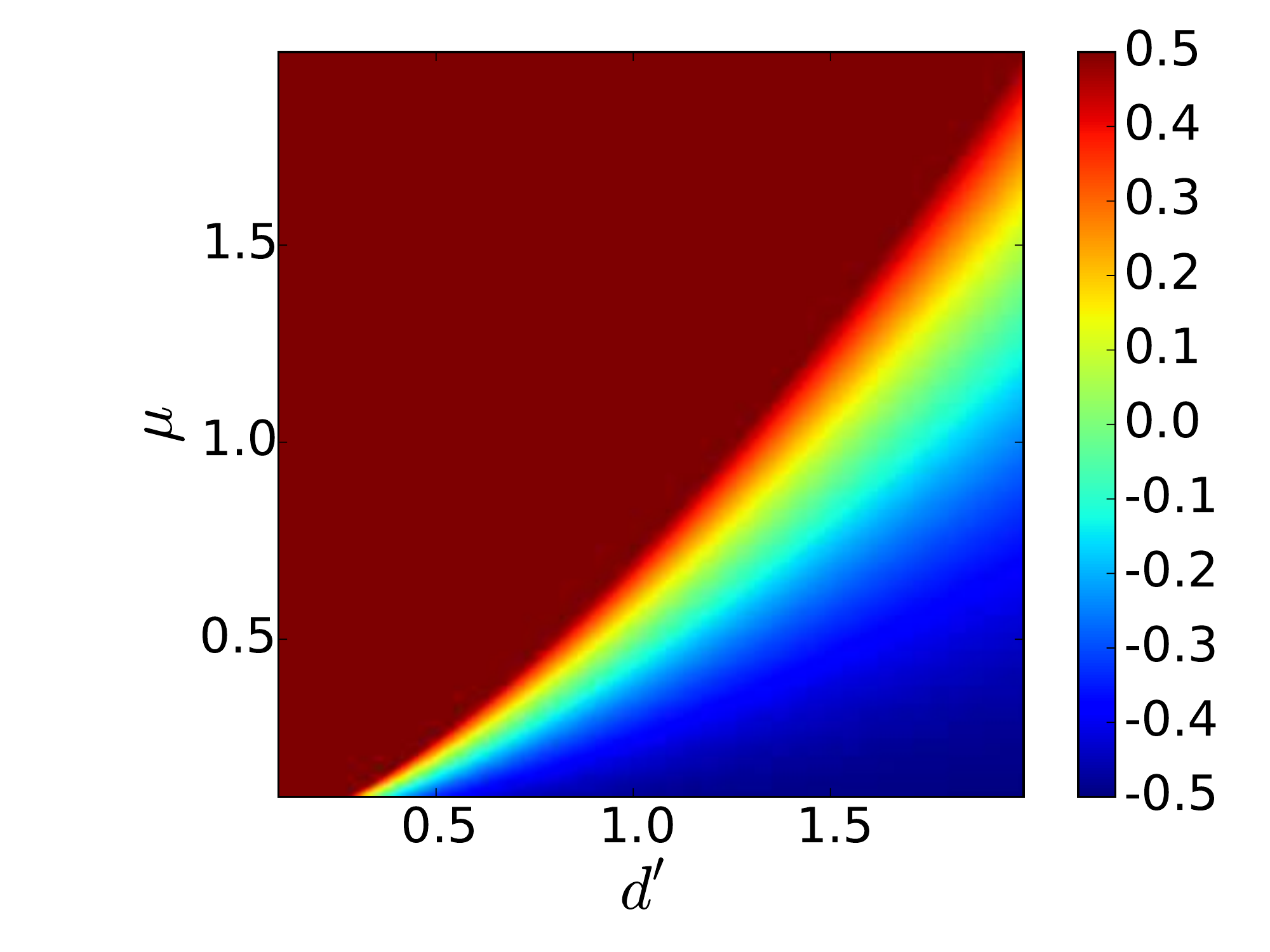}
    \subfigure{}
    \includegraphics[width=0.32\columnwidth]{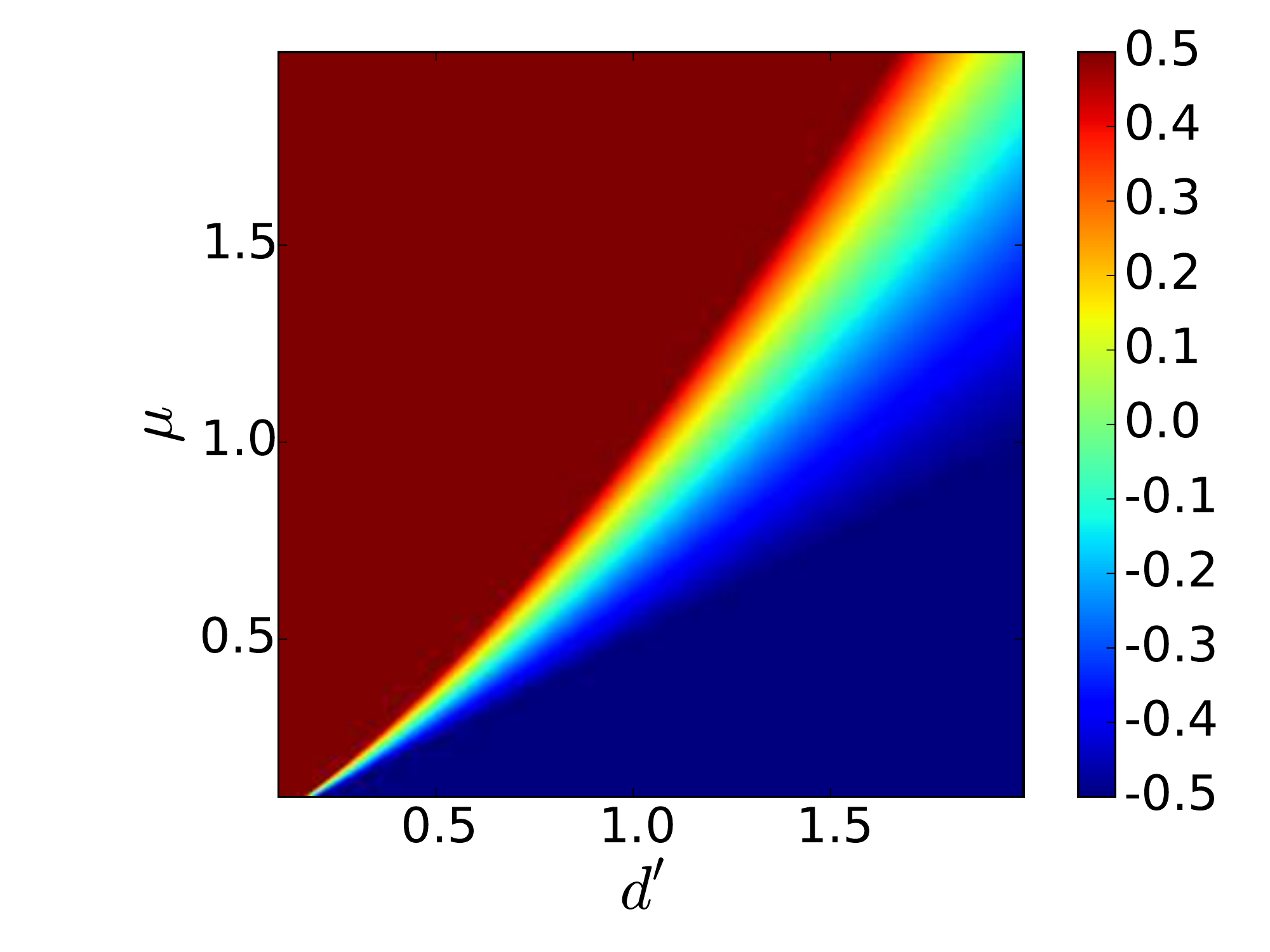}
    \subfigure{}
    \includegraphics[width=0.32\columnwidth]{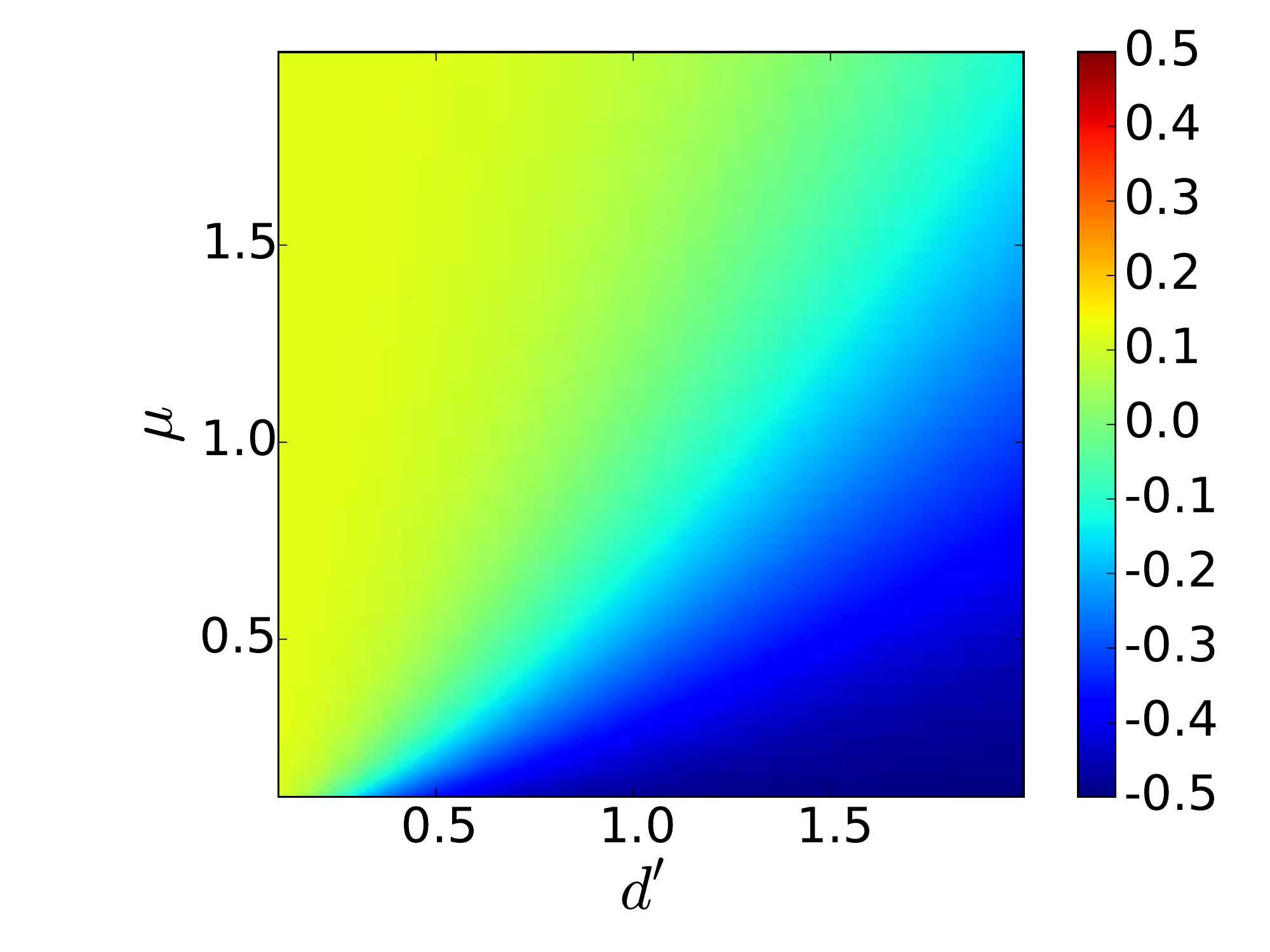}
\vspace{-0.4cm}
    \caption{(Color online) Stability region of the algebraically decaying models with
      respect to $\mu$ and $d'$.
      Left: $\mathcal{Q}=\langle \mu , 0,  2, 0\rangle$.
      Middle: $\mathcal{Q}=\langle \mu , 1, 2, 0\rangle$.  Right:
      $\mathcal{Q}=\langle \mu , 0, 2, 0.2\rangle$.  The colors are
      mapped to the value of $\Phi$ in Eq.~(\ref{condition1}).
      Negative values of $\Phi$ indicate stability regions.}
    \label{fig:rec-models}
  \end{center}
\end{figure}

The stability condition (\ref{condition1}) suggests that models of
type $\mathcal{Q}=\langle \mu\ne 0, 0, q, 0\rangle$, e.g.\ the HFV, GUO
models, tend to instability with increasing density and increasing
strength of the force ($\mu$), because $\Phi$ simplifies to
$\phi-\frac{1}{2}$.  
Adding the influence of the relative speed ($\delta\ne 0$) leads to a
comparable structure (compare Fig.~\ref{fig:rec-models} left and middle).

Modifying these models by introducing a velocity-dependent enlargement
of pedestrians i.e.\ considering models in class $\mathcal{Q}=\langle
\mu\ne 0, 0, q, \tilde a_v\ne 0\rangle$, leads to $\Phi = \phi\omega
-\frac{1}{2}$, with smaller $\omega$ by increasing $\tilde a_v$, which
has a stabilizing effect on the system (see Fig.~\ref{fig:rec-models} right).  
This means the velocity-dependence in this
kind of models enhances the stability of the system. In comparison,
the impact of the relative velocity on the stability of the system is
less significant.

Inverting the sign of $\delta$ adds a positive term to $\phi\omega^2$
in the expression of $\Phi$, which increases the instability of the system
Although negative values of $\delta$ give rise to instabilities, they
are physically not relevant, since that would imply that a faster
pedestrian in front has more influence on a slower pedestrian directly
behind.

\subsection{Simulations}
\label{sec:sim}

We solve the system of equations (\ref{eq:invd}) for $N=67$
using Heun's scheme with time step $\Delta t=10^{-5}$~s. 
According to \cite{Treiber2015} Heun's  scheme seems to be the best 
scheme for  simulations of pedestrian dynamics for many practical scenarios.
For all simulations performed in this work we use this scheme with an unchanged $\Delta t$.

Pedestrians are uniformly distributed in a one-dimensional system with
periodic boundary conditions and length $L=200$~m. The chosen 
values of $N$ and $L$ lead to $d'\approx 1$ ($\tilde a_v=0$). $v_0'=3$.
The initial velocities are set to zero. The maximum simulation time is
$\Delta t = 2000$~s. 
Only the initial position of the first pedestrian is slightly perturbed,
i.e. $\epsilon_1=10^{-4}\,$ ($\epsilon_{n\ne 1}=0$).

With $\Phi=0$ in Eq.~(\ref{condition}) we obtain for $\delta=\tilde
a_v=0$ the critical value for $\mu$ as
$\mu_{\rm{cr}}=\sqrt{\frac{d'^{q+1}}{2q}}$.  Therefore, a model of
type $\mathcal{Q}=\langle 0.45, 0, 2, 0\rangle$ is stable since
$\mu=0.45$ is smaller than this critical value
$\mu_{\rm{cr}}=\frac{1}{2}$.


To observe the behavior of the system in the unstable regime we
perform simulations for a parameter set $\mathcal{Q}=\langle 0.55, 0,
2, 0\rangle$ with $\mu > \mu_{\rm cr}$. The simulations show an
oscillatory behavior that leads inevitably to overlapping among
pedestrians. Note that the model is not
defined when the distance $d'$ is zero, see Eq.~(\ref{eq:invd}). 
This phenomenon (overlapping) is a stopping criterion for the simulation.

\begin{figure}[H]
  \begin{center}
    \subfigure{}
    \includegraphics[width=0.32\columnwidth]{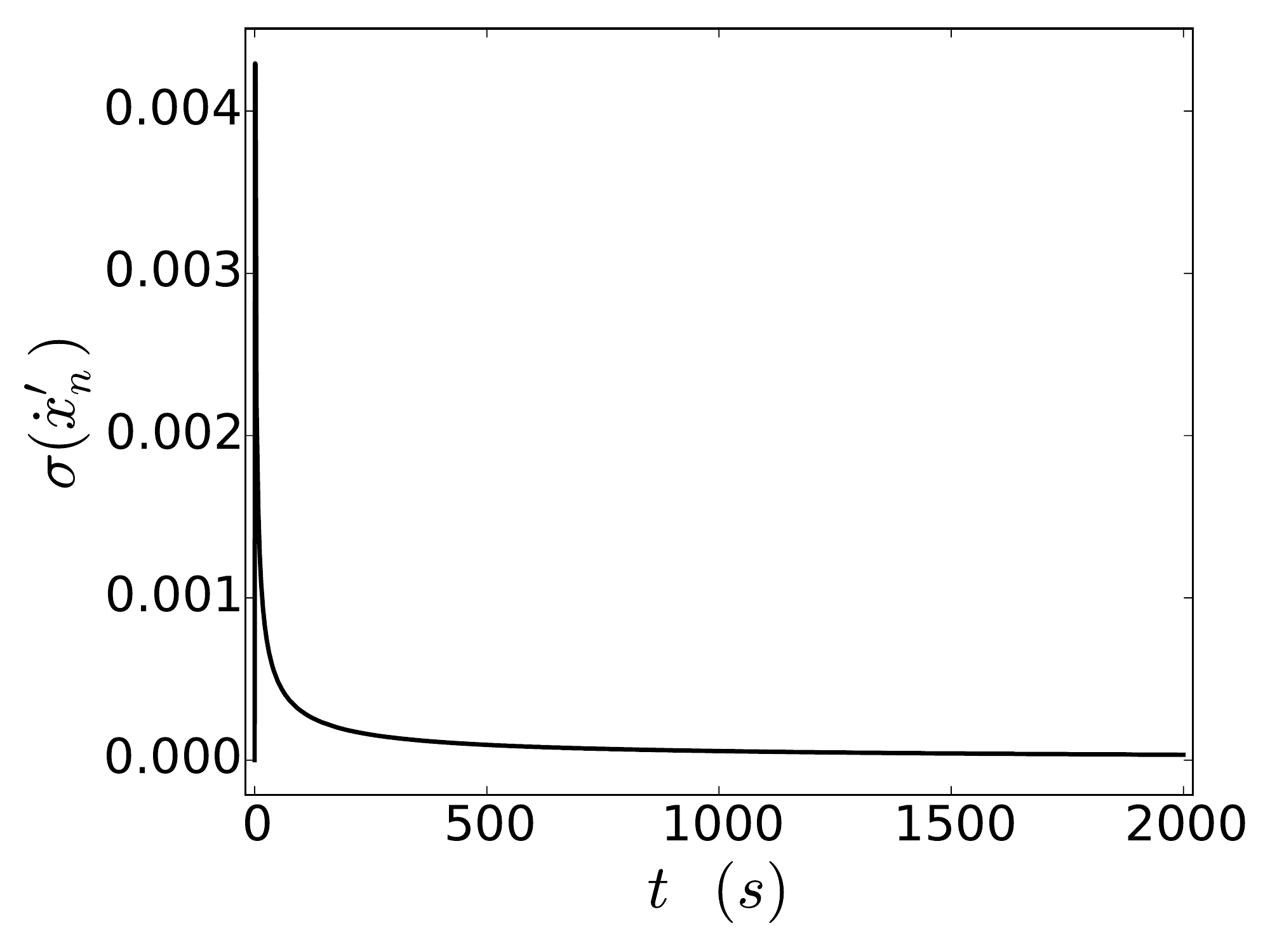}
    \subfigure{}
    \includegraphics[width=0.32\columnwidth]{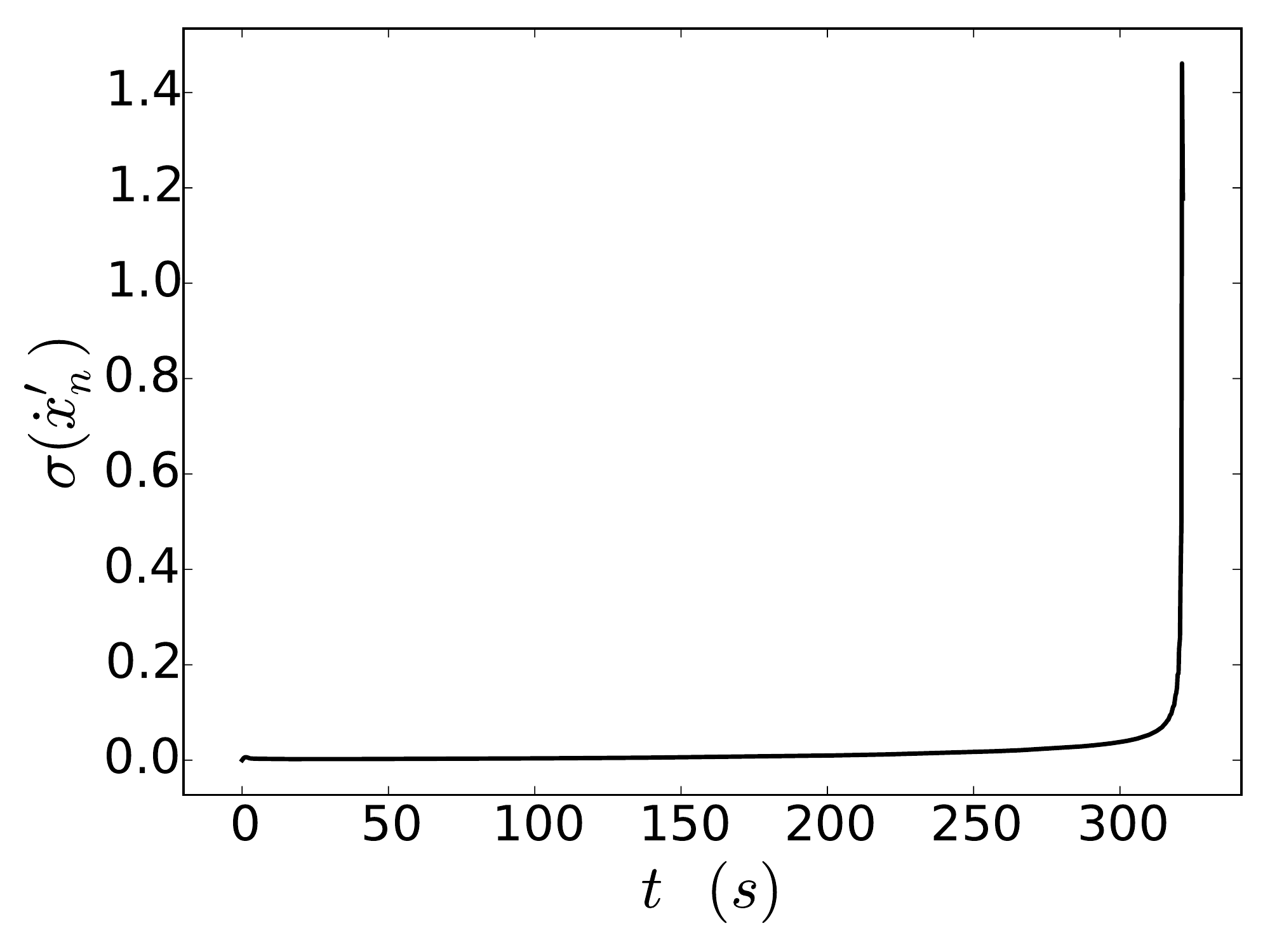}
    \vspace{-0.4cm}
    \caption{Standard deviation of the speeds with respect to simulation time. The initial perturbation
      in the speed disperses to zero when the system is stable (left: $\mu=0.45$), while it
      grows when the system is unstable (right with $\mu=0.55$).}\label{fig:stdv_rec}
  \end{center}
\end{figure}

Since all pedestrians start with speed zero and 
due to the small perturbation of the initial position ($\epsilon_1$) the speeds of pedestrians
in the beginning of simulations are perturbed too. However, depending on the state of the system 
this initial perturbation may disperse to zero if the system is stable. Otherwise, it will grow 
until the simulation is stopped due to overlapping.
Fig.~\ref{fig:stdv_rec} shows a comparison between the time evolution of the speed's standard deviation for both 
cases   
$\mathcal{Q}=\langle 0.45, 0, 2, 0\rangle$ and $\mathcal{Q}=\langle 0.55, 0, 2, 0\rangle$.

We conclude that in the unstable regime the investigated models with
algebraic forces lead to negative velocities (backward movement) and
hence unrealistic behavior. Introducing a velocity-dependent
enlargement of pedestrians stabilizes the system, but the unstable
regime remains unrealistic since the volume exclusion of a pedestrian
($a'_n$) with a negative speed can become negative.  


\section{Exponential-distance models}

In this section we consider models with 
\begin{equation}
  f\Big(\dot x_n,
  \Delta x_n\Big) \propto \exp\Big(- d'_n\Big)\,,
\end{equation}
i.e.~exponentially decaying repulsive forces using use the notation
introduced in the previous section.

The paradigmatic model in this class is arguably the social force
model (SFM) as originally introduced in~\cite{Helbing1995}.  Further
modifications and enhancements followed. In~\cite{Helbing2000} a
physical force was introduced to mitigate overlapping among
pedestrians. Lakoba et al.~\cite{Lakoba2005} studied the calibration
of the modified SFM by improving the numerical efficiency of the model
and introducing several enhancements. The calibration of the modified
SFM was investigated again in~\cite{Johansson2007} by means of an
evolutionary optimization algorithm.  Parisi et al.~\cite{Parisi2009}
investigated the difficulties of SFM concerning quantitative
description of pedestrian dynamics by introducing a mechanism, called
``respect mechanism'' to mitigate overlapping among pedestrians.
Finally in Ref.~\cite{Moussaid2009} an interesting Ansatz to calibrate
the SFM by means of experimental measurements led to a modified
repulsive force that includes the effect of the distance as well as
the angle between two pedestrians. However, these measurements,
basically from experiments with two pedestrians, are extrapolated to a
crowd with several individuals. Hence, it implicitly assumes that the
superposition of forces can be applied.  This hypothesis, however,
lacks experimental evidence in the context of pedestrian dynamics.
Often different specifications of the repulsive force are adopted, in
form of circular or elliptical equipotential lines. However, for a
one-dimensional analysis both specifications are equivalent.  In
comparison to the models with algebraic forces the exponential force
has no singularity at $d^\prime = 0$. Hence it is defined for all
distances and no regularization is required.

\subsection{Linear stability}
\label{sub-stab2}

One common point among the aforementioned models is their
consideration of a ``physical'' force to mitigate overlapping among
pedestrians.  For the stability analysis we therefore consider the
following system using dimensionless variables:
\begin{equation}
  \ddot x_n' =  - a\exp\left(-\frac{d_n'}{b}\right)   -c\,
  r_\varepsilon(d_n')  + v_0'- \dot x_n',
  \label{eq:sfm}
\end{equation}
with $a$, $b$ and $c$ dimensionless positive constants, $d_n'$ 
as defined in (\ref{eq:effDistn}), $v_0'=\frac{v_0\tau}{a_0}$ and
$r_\varepsilon(\cdot)$ the function (\ref{eq:aTheta}).

The general form of these models contains five parameters. However,
the value for $\tau$ was determined empirically in
\cite{Helbing2003,Moussaid2009}. That means the system (\ref{eq:sfm})
can be defined by the quadruple
\begin{equation}
  \mathcal{\tilde Q}=\langle a, b, c, \tilde a_v \rangle.
\end{equation}


Similarly to Sec.~\ref{sub-stab1} we consider the effect of small
perturbations $\epsilon_n(t)=\alpha_n e^{zt}$ to the steady state
positions $y_n$. After some calculations outlined in the appendix
 Sec.~\ref{AppB} we obtain the following stability condition
\begin{equation}
  \tilde \Phi \coloneqq -\frac12+\tilde c \alpha<0,
  \label{eq:stab_condition_exp}
\end{equation}
with $\alpha=\frac1{2\tilde b -1}$, $\tilde b =\tilde a_v \tilde c$, 
$\tilde c=\tilde a/b-\frac{1}{2}c$ and $\tilde a=-a\exp(-d'/b)$.

\begin{figure}[H]
  \begin{center}
    \subfigure{}
    \includegraphics[width=0.32\columnwidth]{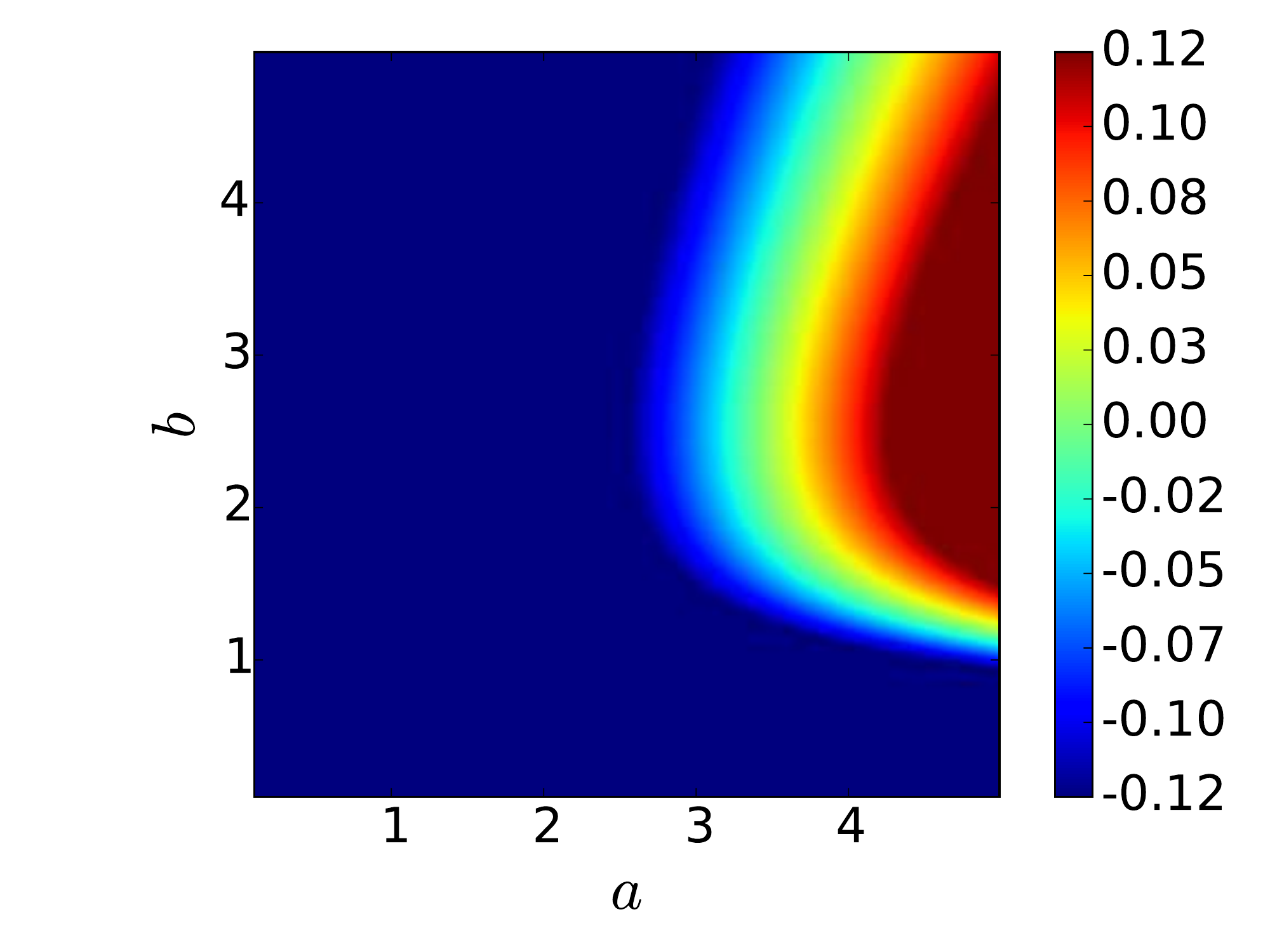}
    \subfigure{}
    \includegraphics[width=0.32\columnwidth]{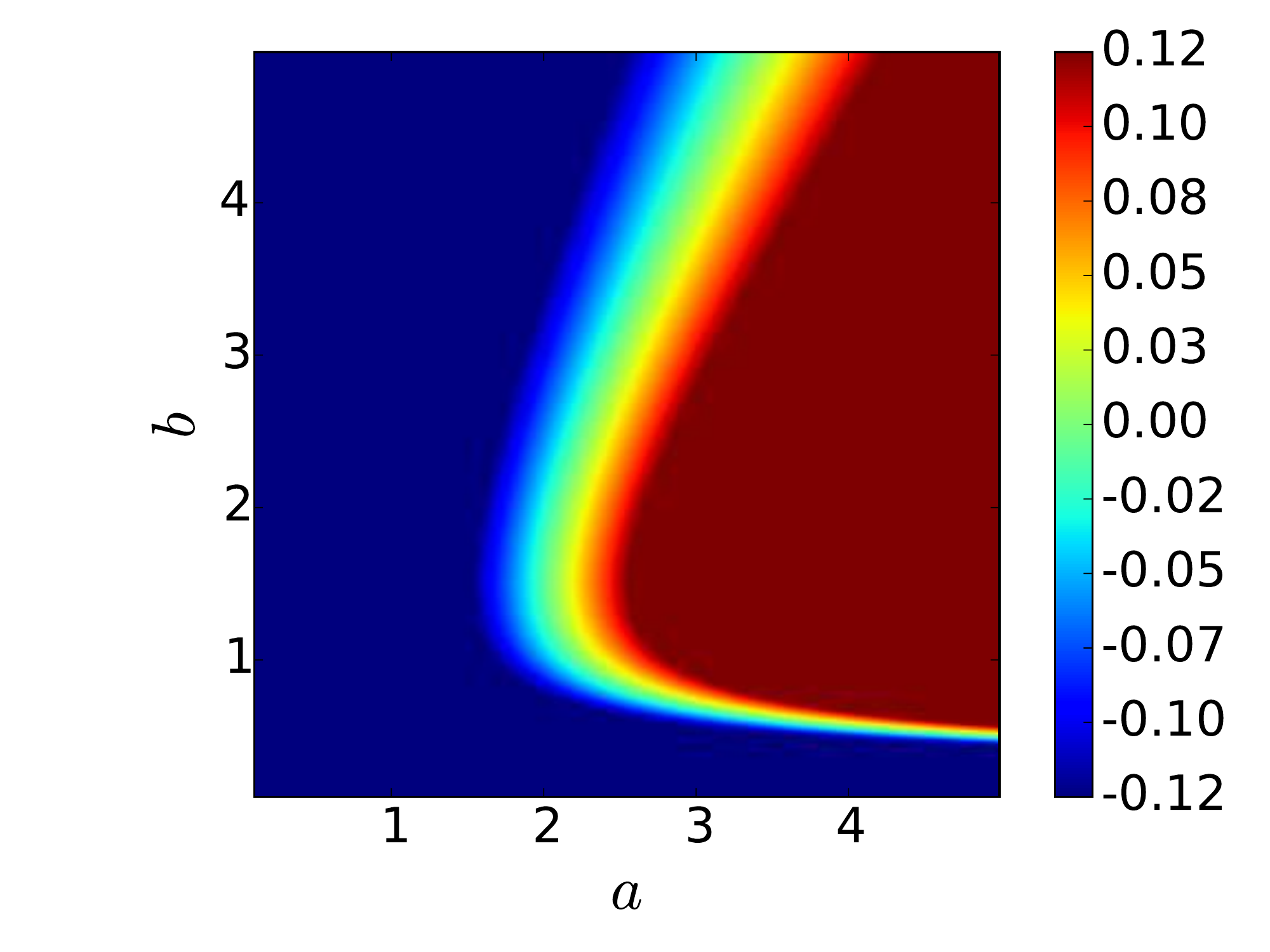}
    \subfigure{}
    \includegraphics[width=0.32\columnwidth]{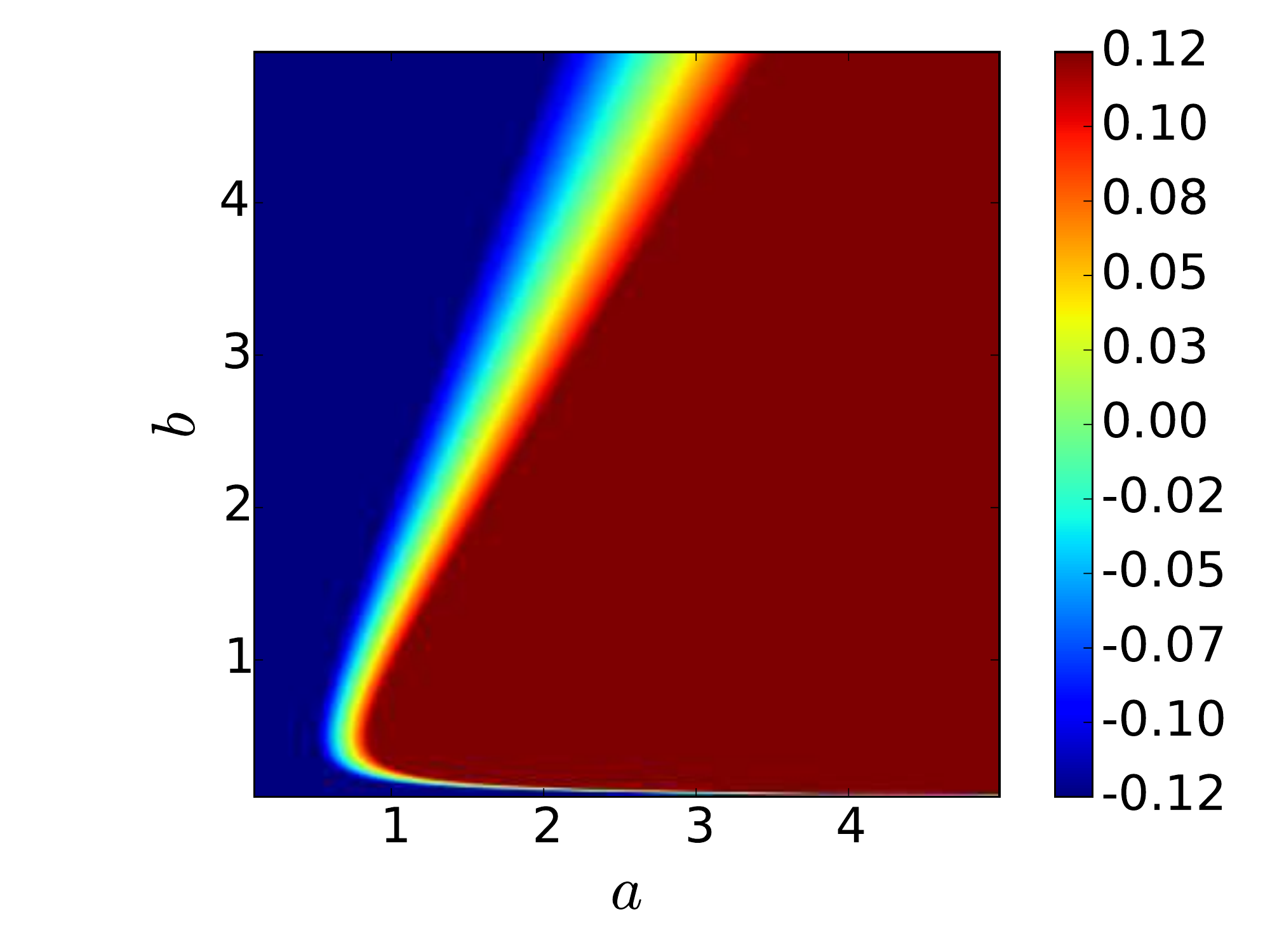}
\vspace{-0.4cm}
    \caption{(Color online) Stability region of a modified SFM ($\mathcal{\tilde Q}=\langle
      a, b, 0, 0\rangle$) with respect to $a$ and $b$ for different
      densities. Left: $d'=2.5\,$. Middle: $d'=1.5\,$. Right: $d'=0.5\,
      $. The colors
      are mapped to the values of $\tilde \Phi$ (Eq. (\ref{eq:exp_stab_sp})). 
      Negative values indicate stability regions.}
    \label{fig:sfm}
  \end{center}
\end{figure}

Assuming $d'$ is positive, which means $r_\varepsilon(\cdot)$ vanishes
or simply $c=0$, and the enlargement of pedestrians is
constant ($\tilde a_v=0$), we obtain 
\begin{equation}
  \tilde b = 0,\;\; \alpha=-1,
\end{equation}
and 
\begin{equation}
  \tilde \Phi = -\frac12+\frac{a}{b}\exp(-d'/b).
  \label{eq:exp_stab_sp}
\end{equation}
Fig.~\ref{fig:sfm} depicts the stability regions for the
$\mathcal{\tilde Q}=\langle a, b, 0, 0\rangle$-class models in the
$(a,\,b)$-plane.

To investigate the effect of a velocity-dependent enlargement of
pedestrians we evaluate the stability regions of $\mathcal{\tilde Q
}=\langle 4, b, 0, \tilde a_v\rangle$-class models. The value of
$a=4$ is according to Fig.~\ref{fig:sfm} large enough to lay in an
unstable region.

\begin{figure}[H]
  \begin{center}
    \subfigure{}
    \includegraphics[width=0.32\columnwidth]{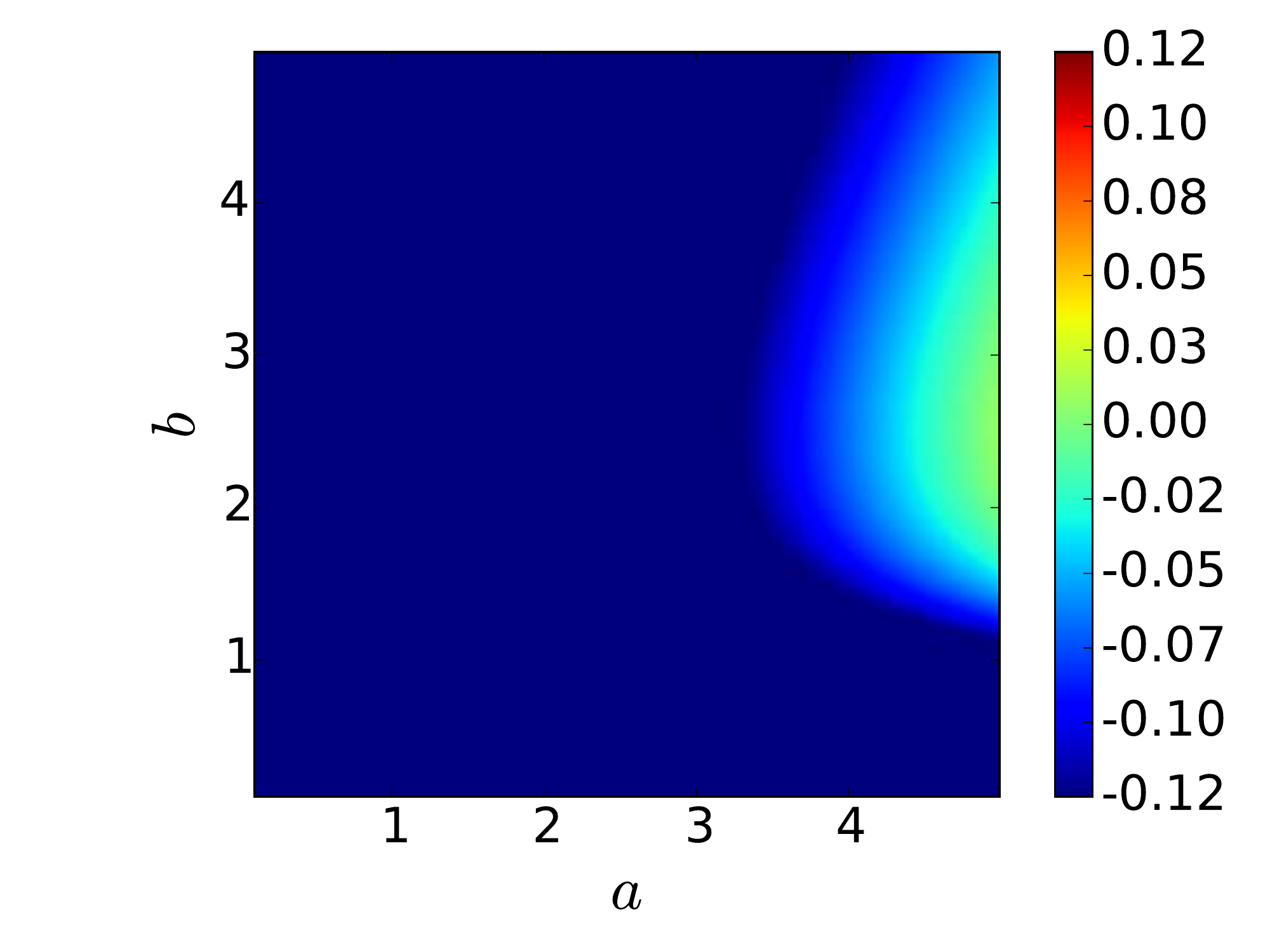}
    \subfigure{}
    \includegraphics[width=0.32\columnwidth]{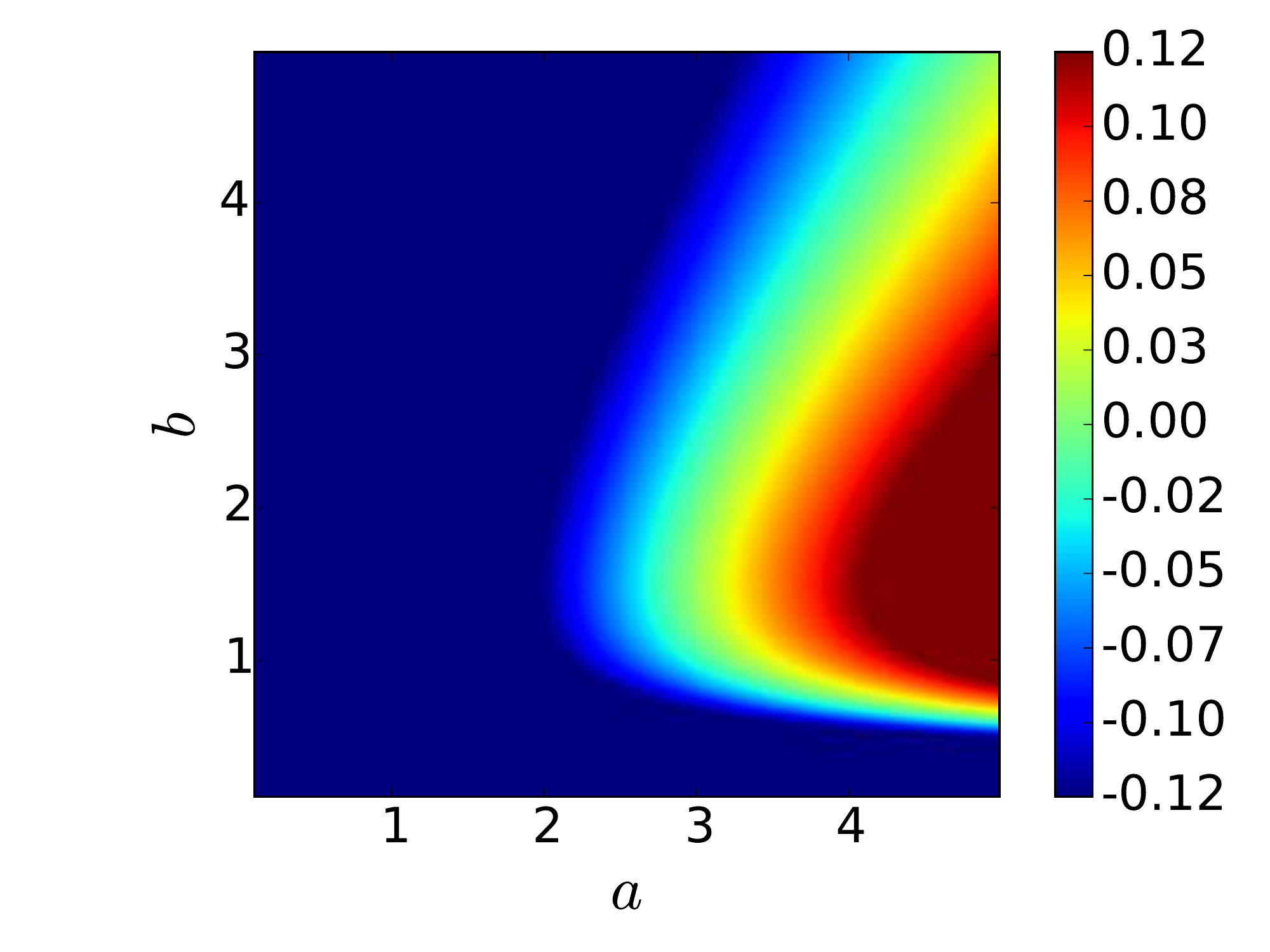}
    \subfigure{}
    \includegraphics[width=0.32\columnwidth]{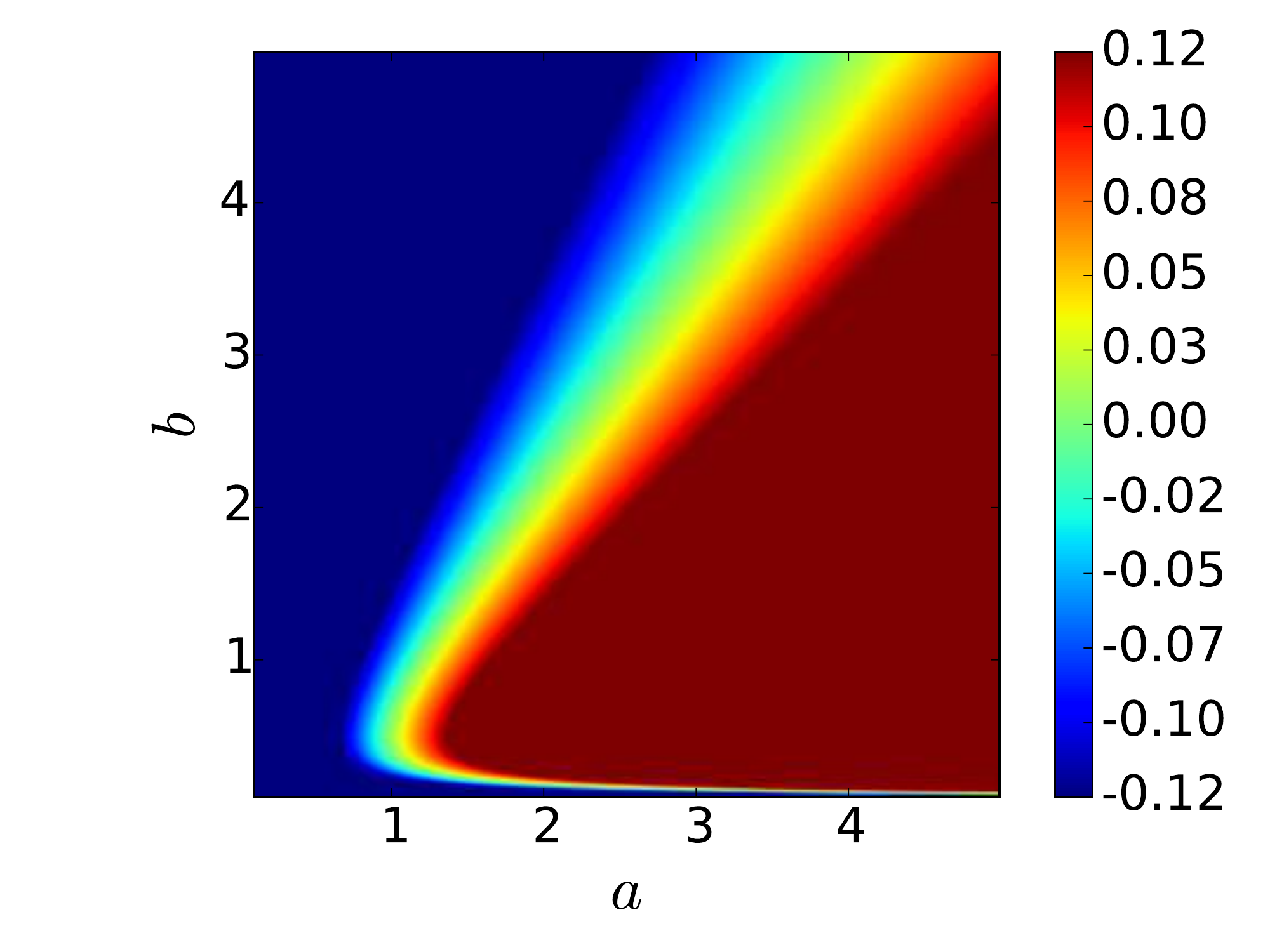}
    \caption{(Color online) Stability region of a modified SFM ($\mathcal{\tilde Q}=\langle
      a, b,  0, \tilde a_v =0.15\rangle$) with respect to $a$ and $b$ for
      different densities. Left: $d'=2.5\,$. Middle:
      $d'=1.5\, $. Right: $d'=0.5\, $.
    }
    \label{fig:avsfm}
  \end{center}
\end{figure}

In Fig.~\ref{fig:avsfm} we observe that a system with a
velocity-dependent enlargement ($\tilde a_v \ne 0$) becomes
increasingly stable in the $(\tilde a, b)$-space with decreasing
density. This confirms the observation made in the previous section:
velocity-dependent enlargement of pedestrians has a stabilizing effect
on the system.

\subsection{Simulations}
\label{sec:sim2}

Similar to Sec.~\ref{sec:sim} we perform simulations with the
exponential-distance models for different parameters. 
The same initial values and parameters as in Sec.~\ref{sec:sim} are considered.
$N=57$ pedestrians are uniformly distributed, which corresponds to  $d'\approx 1.5$.

For $a_v=c=0$, the critical value of $a$ in dependence of $b$ is given
by $a_{\rm cr}= \frac{b}{2\exp(-d'/b)}$. 
Accordingly we choose $b=1.5$ and $a=3$, which yield an unstable system (compare also to Fig.~\ref{fig:sfm}).

Here again we make the same observation as with algebraically
decaying models (Sec.~\ref{sec:sim}).  In the unstable regime a
$\mathcal{\tilde Q}=\langle a, b, 0, 0\rangle$ models behave
unrealistically.  Instead of jams, collisions occur

\begin{figure}[H]
  \begin{center}
    \subfigure{}
    \includegraphics[width=0.32\columnwidth]{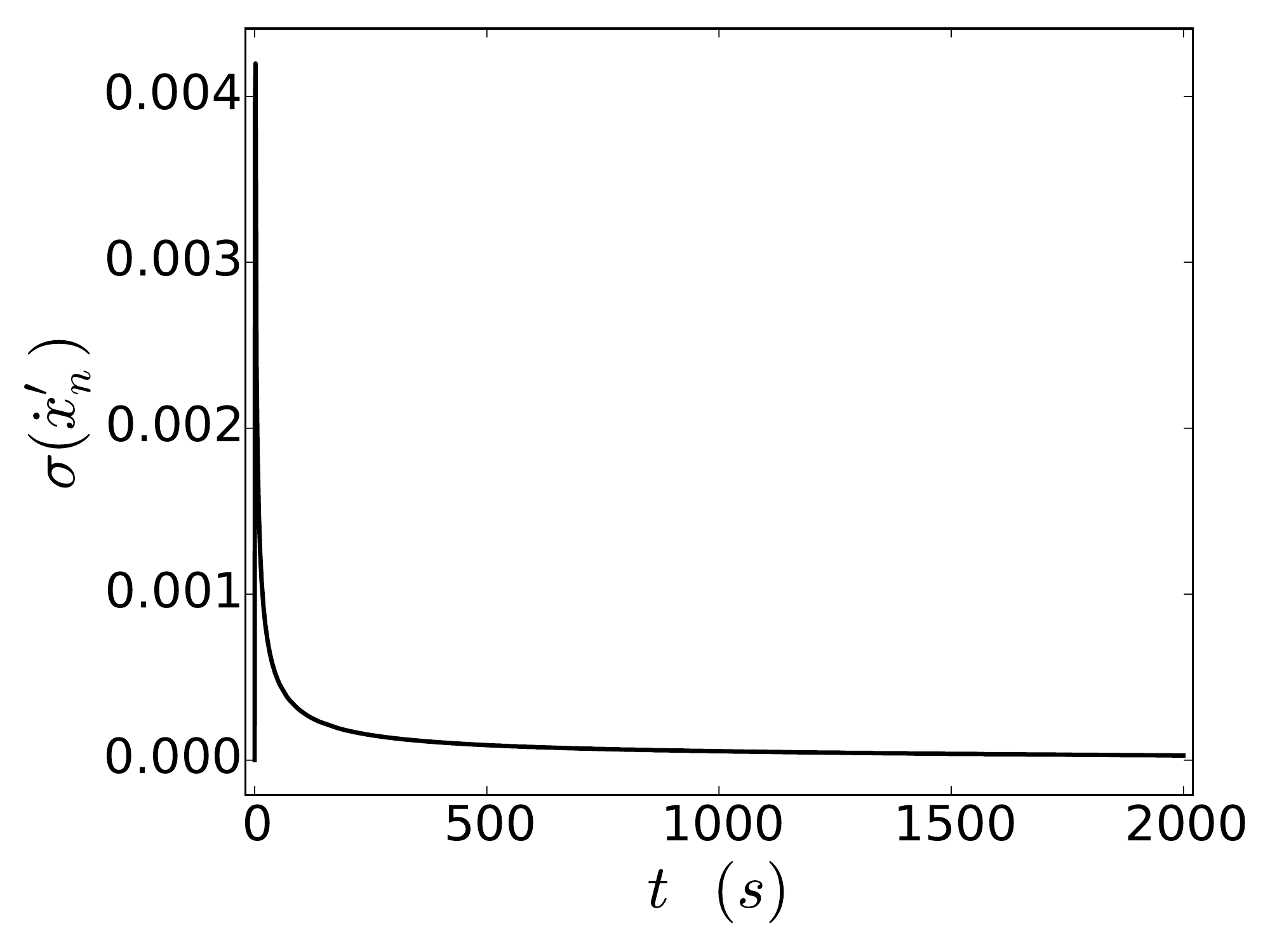}
    \subfigure{}
    \includegraphics[width=0.32\columnwidth]{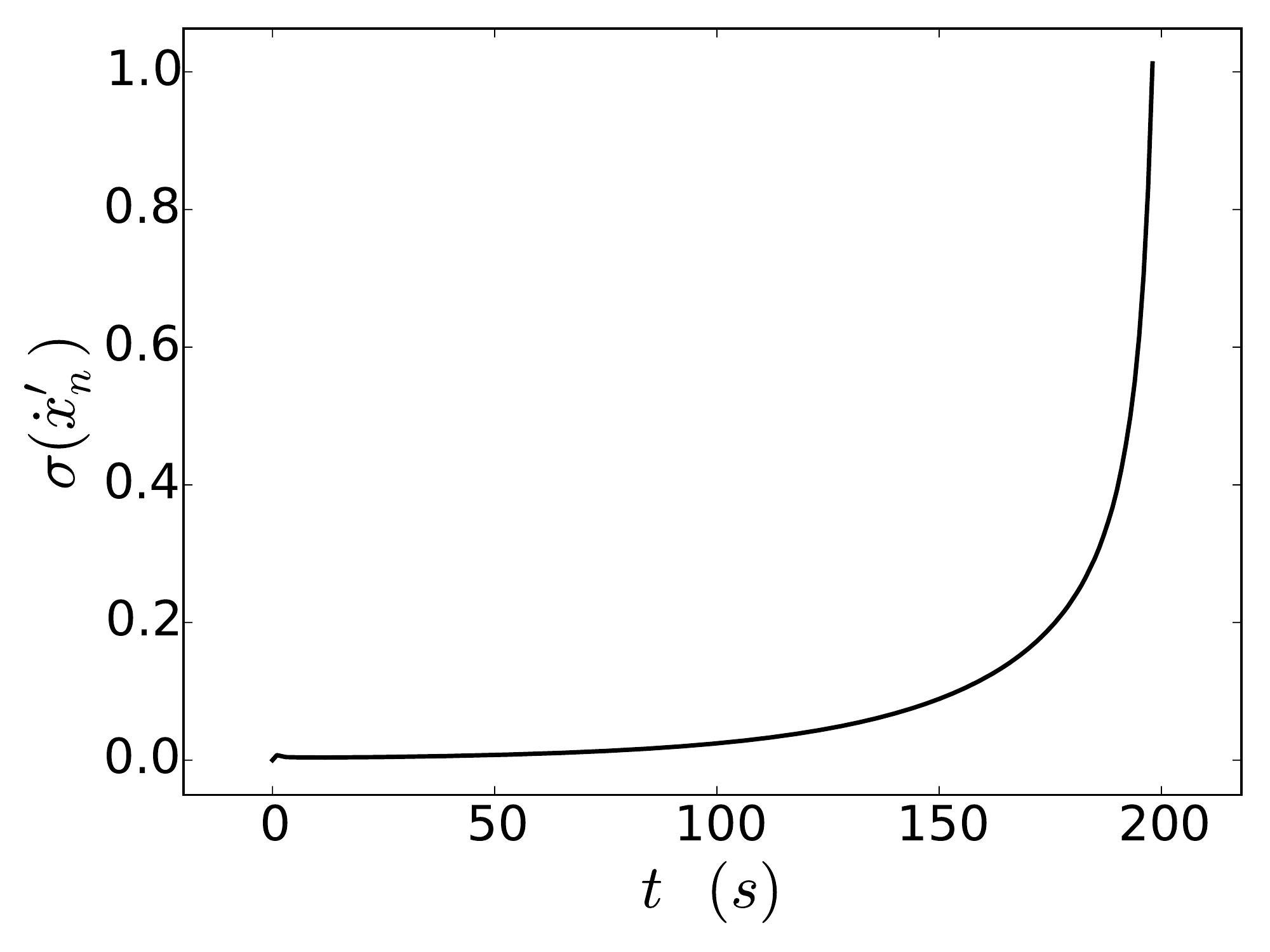}
    \vspace{-0.4cm}
    \caption{Standard deviation of the speeds with respect to simulation time. The initial perturbation
      in the speed disperses to zero when the system is stable (left: $a=1.5,\; b=1.5$), while it
      grows when the system is unstable (right with $a=3.0,\; b=1.5$).}\label{fig:vstd_exp}
  \end{center}
\end{figure}

Based on the time series of the speed's standard deviation, 
we compare the behavior of the model in a stable and an unstable 
regime (defined according to Eq.~(\ref{eq:stab_condition_exp}).
Fig.~\ref{fig:vstd_exp} left shows as expected 
for $\mathcal{\tilde Q}=\langle 1.5, 1.5, 0, 0\rangle$ that the standard deviation of the speed, 
decreases to zero and the overall system converges to an homogeneous state, whereas it grows until the simulation interruption ($\mathcal{\tilde Q}=\langle 3, 1.5, 0, 0\rangle$).

\section{A new model}

In the previous sections we investigated properties of several
force-based models related to jam formation.  The linear stability
analysis of these models yields conditions that determine parameter
regions where unstable behavior may lead to stop-and-go waves in
one-dimensional systems with boundary conditions. However, simulations
with parameters in the unstable regime lead to unrealistic behavior
(collisions, overlapping etc.) instead of stop-and-go waves. 
In this section we discuss the reasons for this failure and formulate a
new model that produces stop-and-go waves in its unstable regime.   


Rewriting the generic equation of motion (\ref{eq1}) as
\begin{equation}
  \ddot x_n = \frac{\tilde v_0-\dot x_n}{\tau},
  \label{eq:modified2}
\end{equation}
with $\tilde v_0 = \tau f+v_0 \le v_0$ implies that the movement of
pedestrian $n$ is determined by a driving force with a modified and
density-dependent desired speed $\tilde v_0$: the higher the density,
the smaller the desired speed.  However, if the desired speed is
negative, which means pedestrians move backwards after some delay, collisions are
likely to happen.  This is in fact the case in the reciprocal-distance
and exponential-distance models, where collisions are observed in the
unstable regimes instead of jams.

In order to avoid such problems, a non-linear function $f(\Delta
  x_n, \dot x_n, \dot x_{n+1})$ such that $f(0, 0, 0) = -v_0/\tau$ is
  required. That means that overlapping of pedestrians leads
to a vanishing desired speed $\tilde v_0=0$ instead of a negative one.
Note that initial high values of $v_0$ may still lead to backward movement even if the resulting 
desired speed  $\tilde v_0=0$. We discuss this effect in more detail in section \ref{sec:discussion}.

\begin{figure}[H]
  \begin{center}
    \includegraphics[width=0.32\columnwidth]{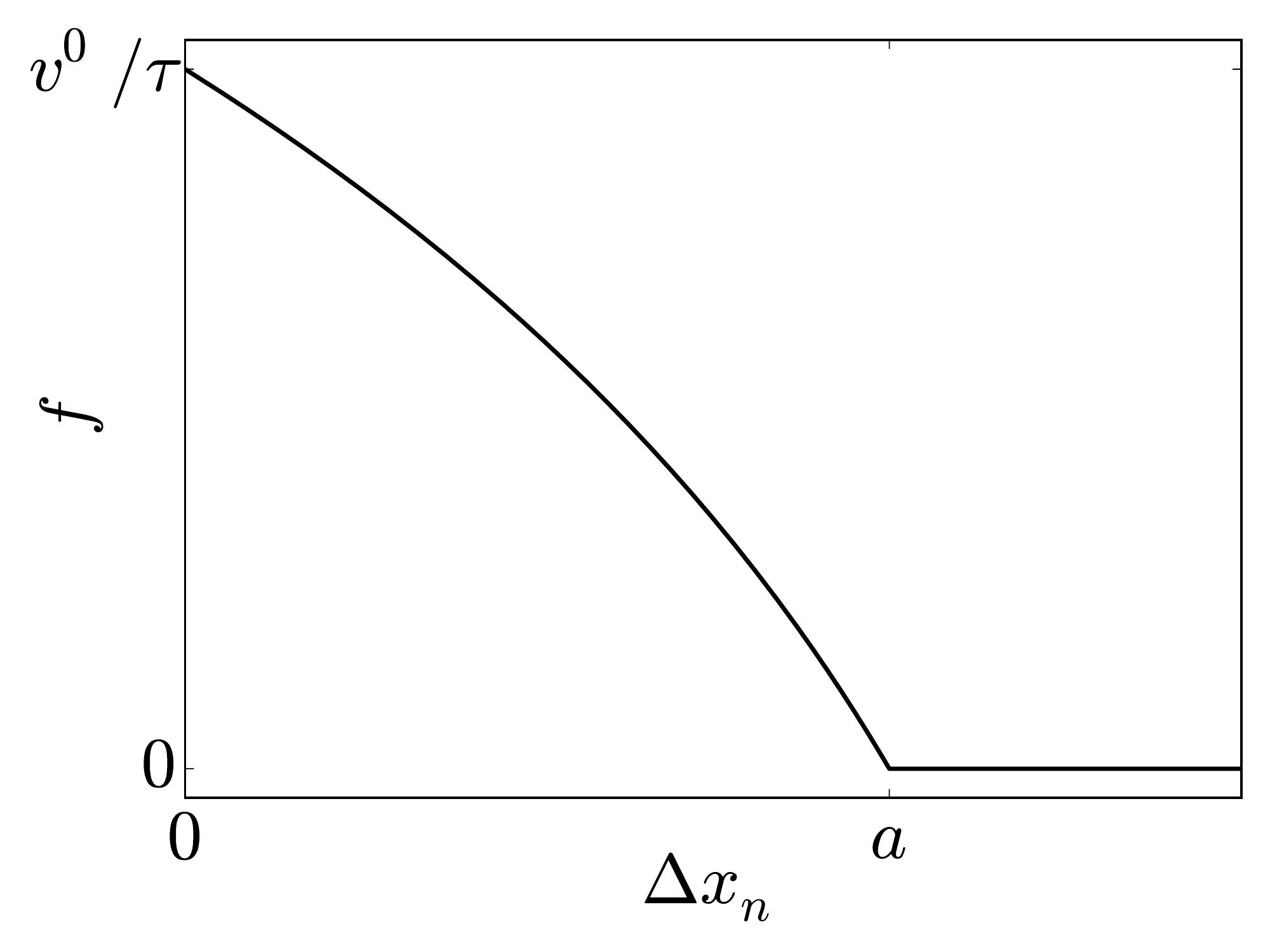}
\vspace{-0.4cm}
    \caption{The absolute value of the repulsive force according to Eq. (\ref{eq:newf}).}
    \label{fig:log_f}
  \end{center}
\end{figure}

For $f$ we propose the following expression:
\begin{equation}
  f(\Delta x_n, \dot x_n, \dot x_{n+1}) = - \frac{v_0}{\tau}
  \log\Big(c\cdot R_n + 1\Big),
  \label{eq:newf}
\end{equation}
with 
\begin{equation}
  \qquad R_n =r_\varepsilon 
  \Big(\frac{\Delta x_n}{a_n+a_{n+1}} -1\Big),\qquad c = e - 1.
\label{eq:Rn}
\end{equation}
Pedestrians anticipate collisions when their distance to their
predecessors is smaller than a critical distance $a=a_n + a_{n+1}$,
which is given by the addition of safety distances of two consecutive
pedestrians. It is worth pointing out at this point that $a_n$ does
not model the body of pedestrian $n$ but represents a ``personal''
safety distance.  For $\Delta x_n=0$, i.e., $R_n=1$ the repulsive
force reaches the value $-v_0/\tau$ to nullify the effects of the
driving term (Fig.~\ref{fig:log_f}). In other words, the desired speed
$\tilde v_0$ vanishes and pedestrians are not pushed to move
\textit{backwards}.

The corresponding dimensionless model we henceforth use is
\begin{equation}
  \ddot x_n' = -v'_0\ln\Big(c\cdot R_n' + 1\Big)- \dot x_n' + v'_0,
  \label{eq:modlog}
\end{equation}
with 
\begin{equation}
  \quad R_n' 
  =r_\varepsilon \Big(\frac{\Delta x_n'}{a'_n+a'_{n+1}} -1\Big),\quad v'_0
  =\frac{v_0\tau}{a_0}.
\end{equation}
The main difference between this model and the optimal velocity model
\cite{Bando1995,Nakayama2005} is the velocity-dependent space
requirement of pedestrians, expressed by the critical distance $a$.


\subsection{Stability analysis}

\begin{figure}[H]
  \begin{center}
    \subfigure{}
    \includegraphics[width=0.32\columnwidth]{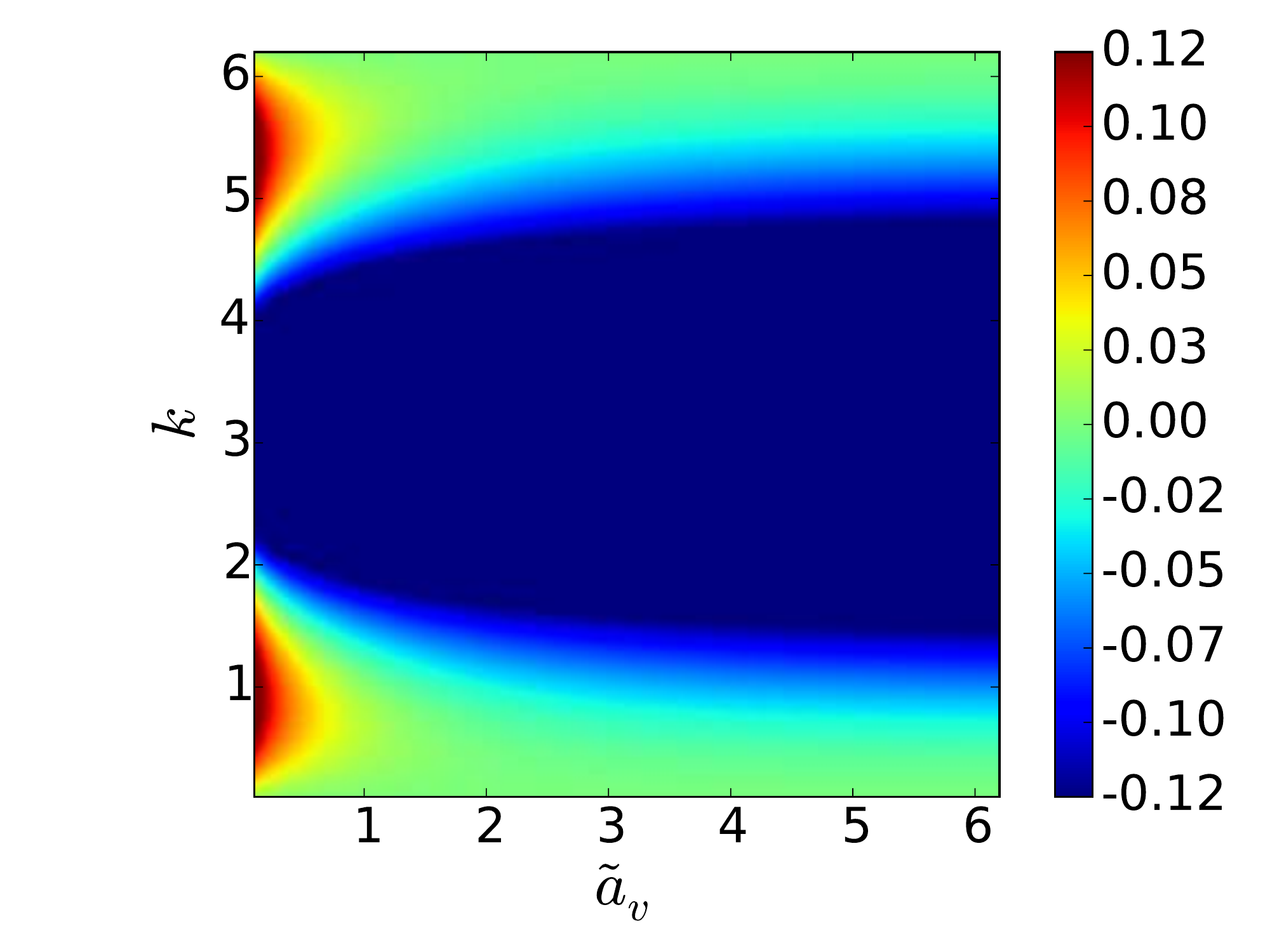}
    \subfigure{}
    \includegraphics[width=0.32\columnwidth]{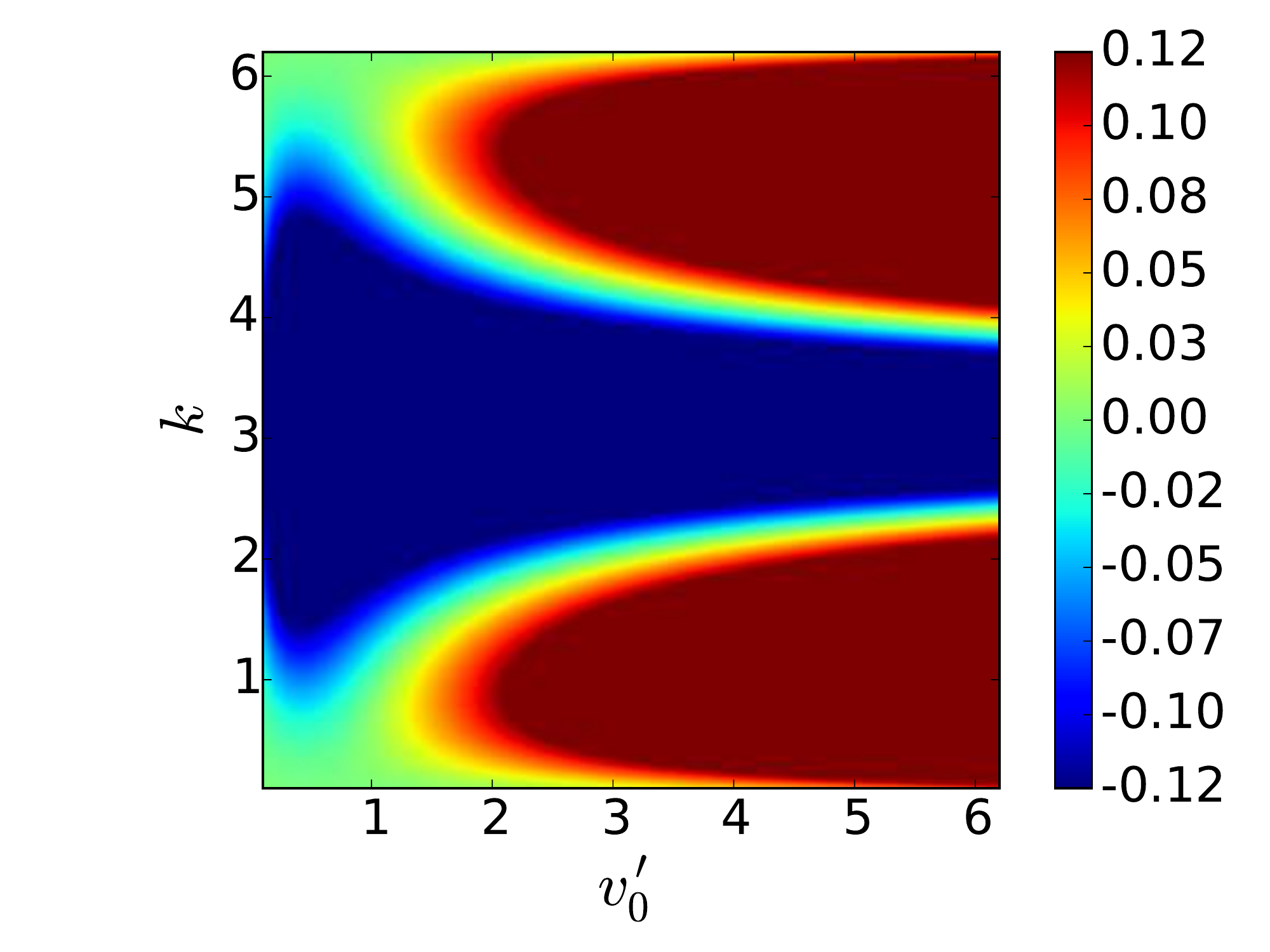}
\vspace{-0.4cm}    \caption{(Color online) Left: Stability region in the $(\tilde a_v, k)$-space for
      $v^\prime_0=3$ and $\Delta y=1.5$. Right: Stability region in the
      $(\tilde v'_0, k)$-space for $\tilde a_v=0$ and $\Delta y=1.5$.
      The colors are mapped to the values of real part of the positive solutions $\hat z^+$.
    }
    \label{fig:newd_av}
  \end{center}
\end{figure}

In this section, we investigate the stability of the new model.  We
suppose that $\Delta y<a'$, with $\Delta y=\frac1{\rho a_0}$ is the
mean dimensionless spacing and $a'=2(1+\tilde a _v v')$, $v'$ being
the dimensionless speed for the equilibrium of uniform solution, and
add a small perturbation $\epsilon_n$ to the dimensionless coordinates
of pedestrians. For $R_n'$ we obtain with $a^\prime =2(1 +\tilde
a_vv')$ and $a_v^\prime = \frac{\tilde a_v}{a^\prime}$
\begin{align}
  R_n' 
         &\approx 1 - \frac{\Delta x'_n}{a^\prime}\Big( 1 - a_v^\prime(\dot \epsilon_n
           + \dot \epsilon_{n+1}) \Big).\;\;
\end{align}
From the equation of motion (\ref{eq:modlog}) we obtain with $d_0=1 +
c\Big( 1 - \frac{\Delta y}{a^\prime}\Big)$
\begin{align*}
  \ln( c\cdot R_n'+1) 
                        &\approx \ln(d_0)  +\frac{c}{d_0}\Big(
                          \frac{\Delta y}{a^\prime}a_v^\prime(\dot \epsilon_n+ \dot
                          \epsilon_{n+1}) -\frac{\Delta \epsilon_n}{a^\prime} \Big).
\end{align*}
Eq. (\ref{eq:modlog}) in steady state yields $v_0'\ln(d_0)  = v_0' - v'$ thus



\begin{equation}
  \ddot \epsilon_n
  =  -v'_0 \frac{c}{d_0}\Big(
  \frac{\Delta y}{a^\prime}a_v^\prime(\dot \epsilon_n+ \dot
  \epsilon_{n+1}) -\frac{\Delta \epsilon_n}{a^\prime} \Big) - \dot \epsilon_n.
  \label{eq:eps_neu}
\end{equation}
Eq. (\ref{eq:eps_neu}) rewritten in the $z$-domain yields
\begin{equation}
  z^2 +\Big(  \xi a_v'\Delta y\Big(e^{\boldsymbol{i}k}+1\Big)
  + 1\Big)z -
  \xi\Big(e^{\boldsymbol{i}k}-1\Big)   =0.
  \label{eq:stab_neu_av}
\end{equation}
with $\xi= \frac{cv'_0}{a' d_0}$. 
Given $\hat z^\pm$ two solutions of (\ref{eq:stab_neu_av}) we show in
Fig.~\ref{fig:newd_av} the influence of the velocity-dependence of the
safety distance ($\tilde a_v$) and the constant $v'_0$ on the
stability behavior of the model.

As expected we observe that velocity-dependent safety distance has a
stabilizing effect on the model. Unlike the previous models for
$a_v\ne 0$ the model still can show significant unstable behavior.
This observation is important since it has been shown in the context of different force-based models that constant space
requirement of pedestrians is responsible for an unrealistic shape of the
fundamental diagram in single-lane movement~\cite{Seyfried2006,Chraibi2010a} 
Additionally, we observe that increasing $v'_0$ leads to an unstable system.

Expanding Eq.~(\ref{eq:stab_neu_av}) around $k\approx 0$ yields the stability condition
\begin{equation}
  \hat \Phi \coloneqq  \Big(\frac{1}{1 + 2\xi a'_v \Delta
    y}\Big)\Big(\frac{\xi}{1 + 2\xi a'_v \Delta y}  +\xi a'_v\Delta y
  \Big) - 1/2 < 0.
  \label{eq:log_condition}
\end{equation}
For $\tilde a_v=0$ the equation above simplifies to
\begin{equation}
  \xi<1/2,\qquad \xi= \frac{cv'_0}{a' d_0}.
\end{equation}
This result is in agreement with the stability condition
$V'<1/(2\tau)$ given in Ref.~\cite{Bando1995} for the system
\begin{equation}
  \ddot x_n = A(V(\Delta x_n) -\dot x),
\end{equation}
with $A=1/\tau$ and $V(\Delta x_n) = v_0(1-\ln(1+cR))$.

\begin{figure}[H]
  \begin{center}
    \includegraphics[width=0.35\columnwidth]{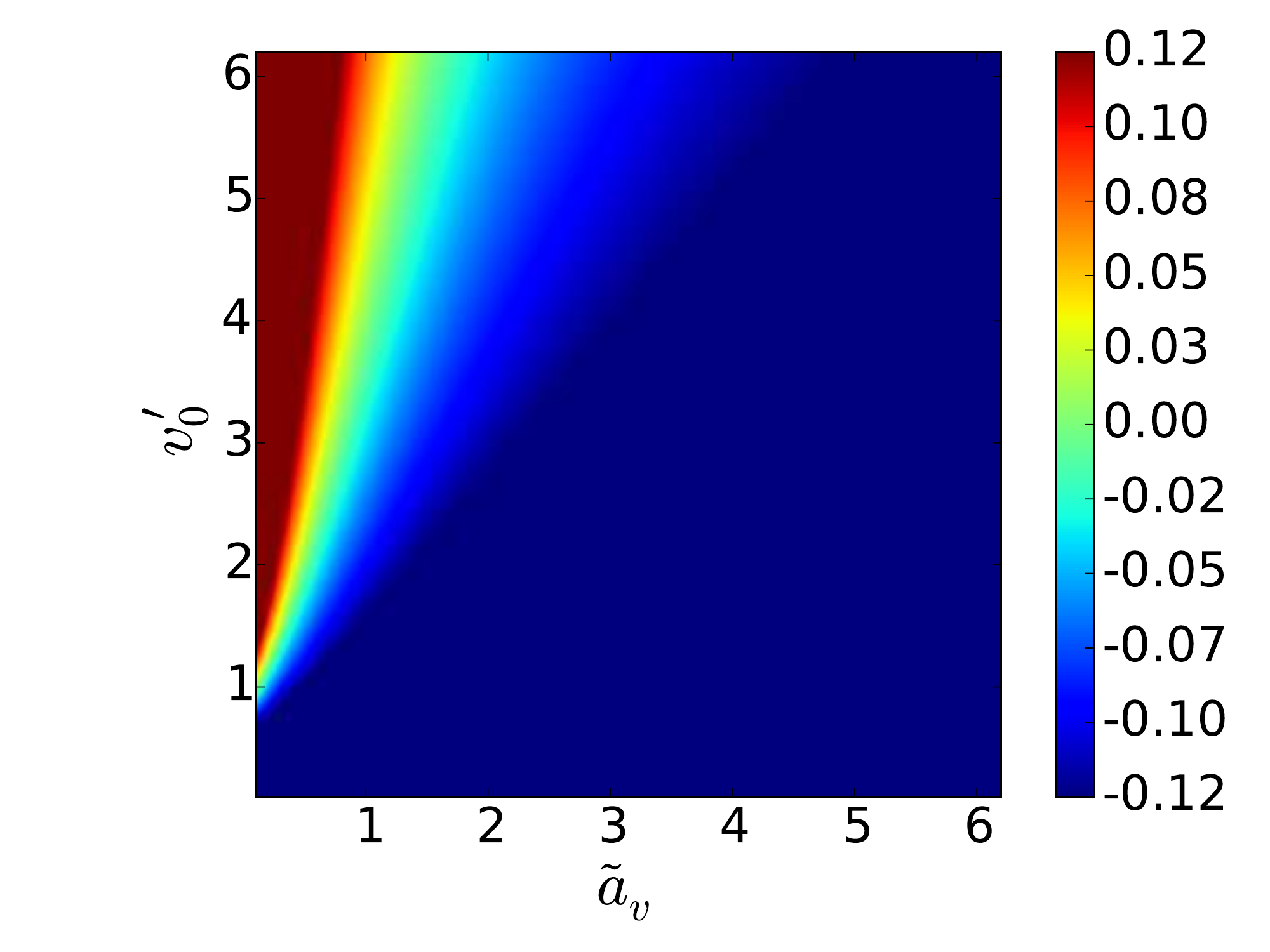}
    \caption{(Color online) Stability region in the $(\tilde a_v, v^\prime_0)$-space for
      $\Delta y=1.5$.  The colors are mapped to the values of $\hat
      \Phi$ in Eq.~(\ref{eq:log_condition}). }
    \label{fig:new_av_v0}
  \end{center}
\end{figure}

The dimensionless from of the equation of motion (\ref{eq:modlog})
has only two free parameters, $v'_0$ and $\tilde a_v$.
In Fig.~\ref{fig:new_av_v0} we observe that the system becomes
increasingly unstable with increasing $v'_0$ (by a relatively small
and constant $\tilde a_v$). Assuming that the free flow speed $v_0$ is
constant, this means that increasing the reaction
time $\tau$ or diminishing the safety space leads to unstable behavior
of the system.

\subsection{Simulations}

\begin{figure}[H]
  \centering
  \includegraphics[scale=0.35]{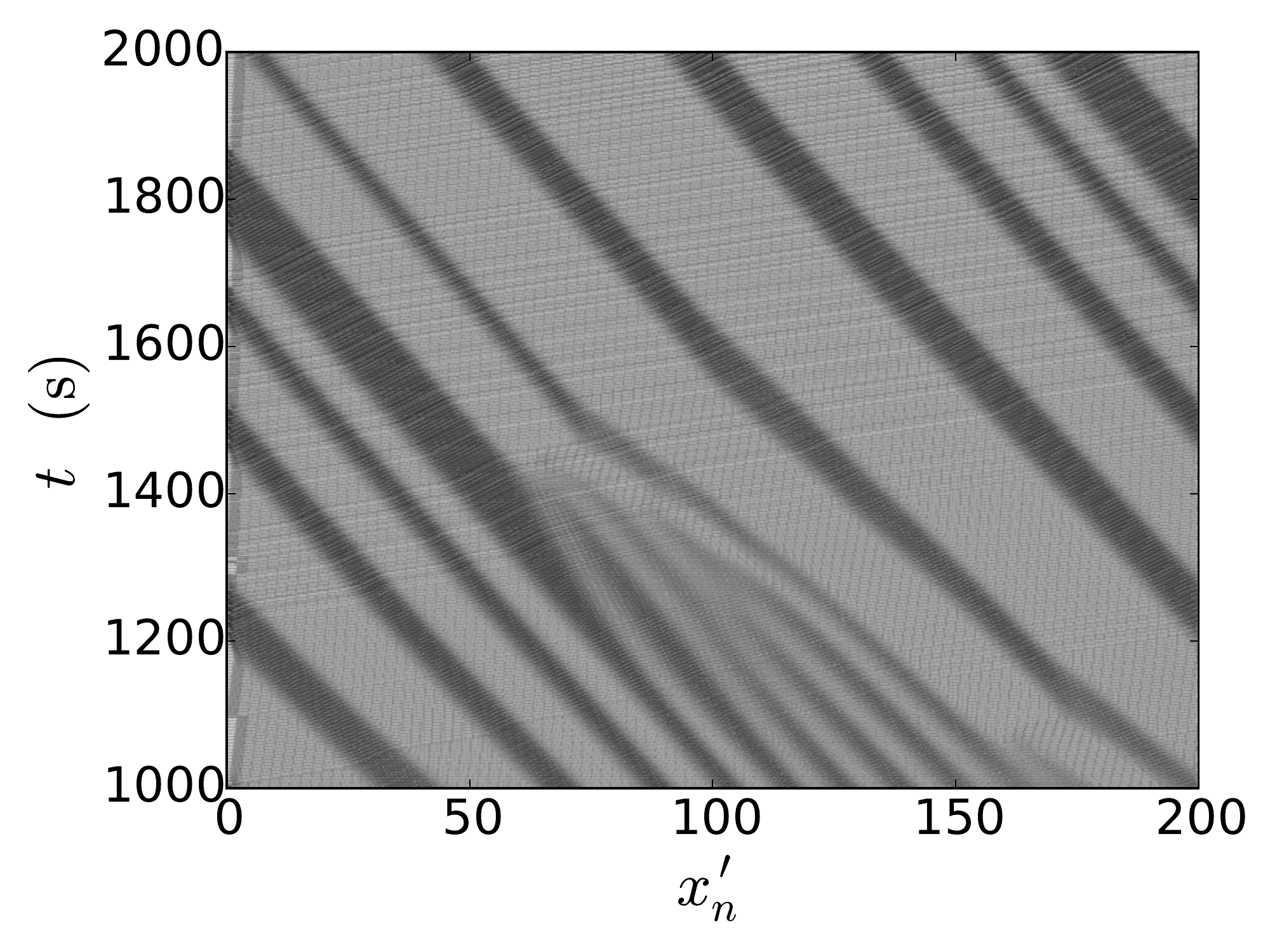}
\vspace{-0.3cm}
  \caption{Trajectories by $\Delta y_n = 1.5$. 
    The trajectories show stop-and-go waves.}
  \label{fig:jams}
\end{figure}

We perform simulations with the introduced models using the same
set-up as before. For $\tilde a_v=0$, $v'_0=1$  and $\Delta
y_n = 1.5$ we calculate the solution for 3000 s. Fig.~\ref{fig:jams}
shows the trajectories of 133 pedestrians. $\varepsilon$ in Eq. (\ref{eq:Rn}) is set to 0.01.

\begin{figure}[H]
  \centering
  \subfigure{}
  \includegraphics[scale=0.32]{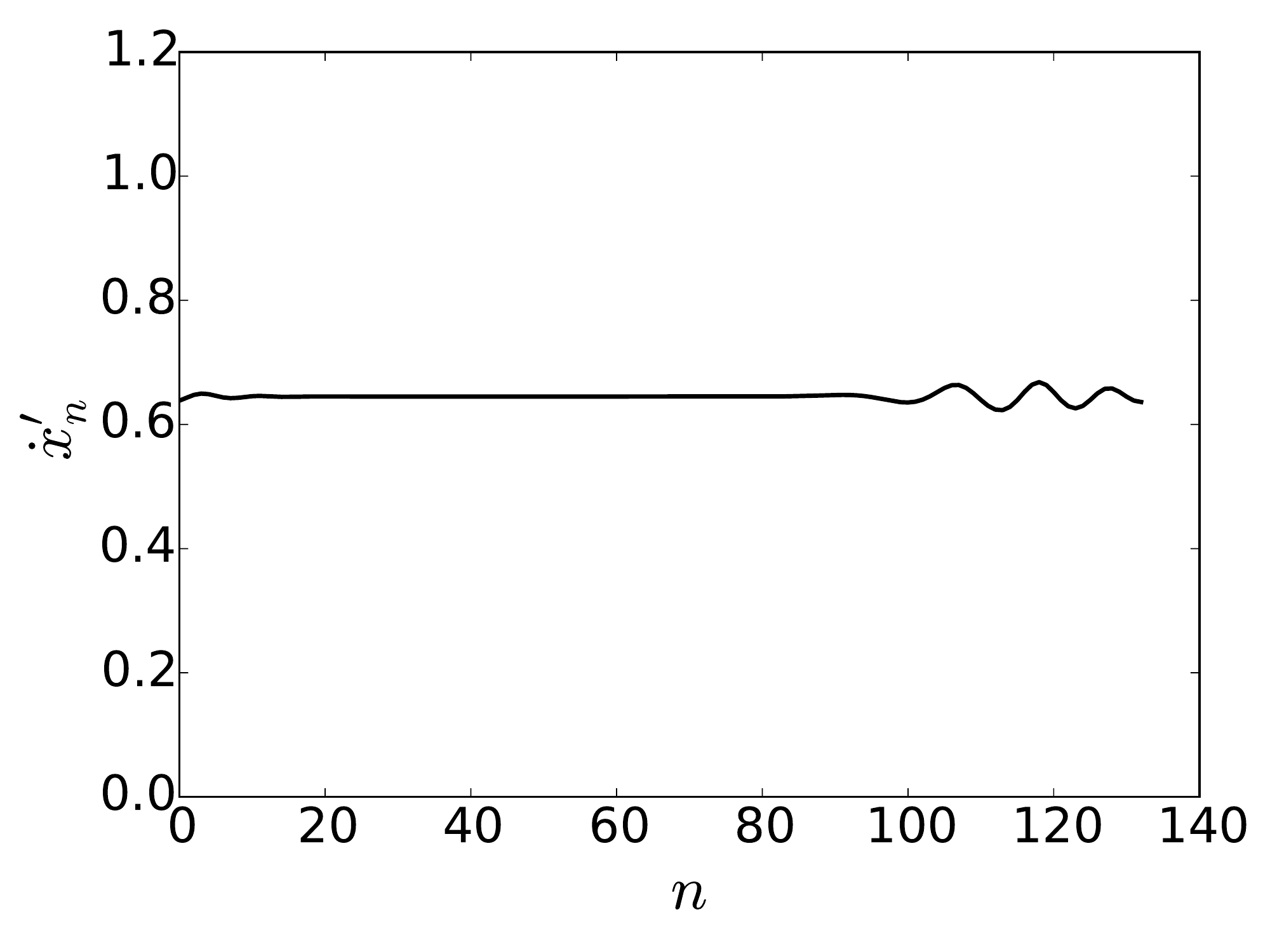}
  \subfigure{}
  \includegraphics[scale=0.32]{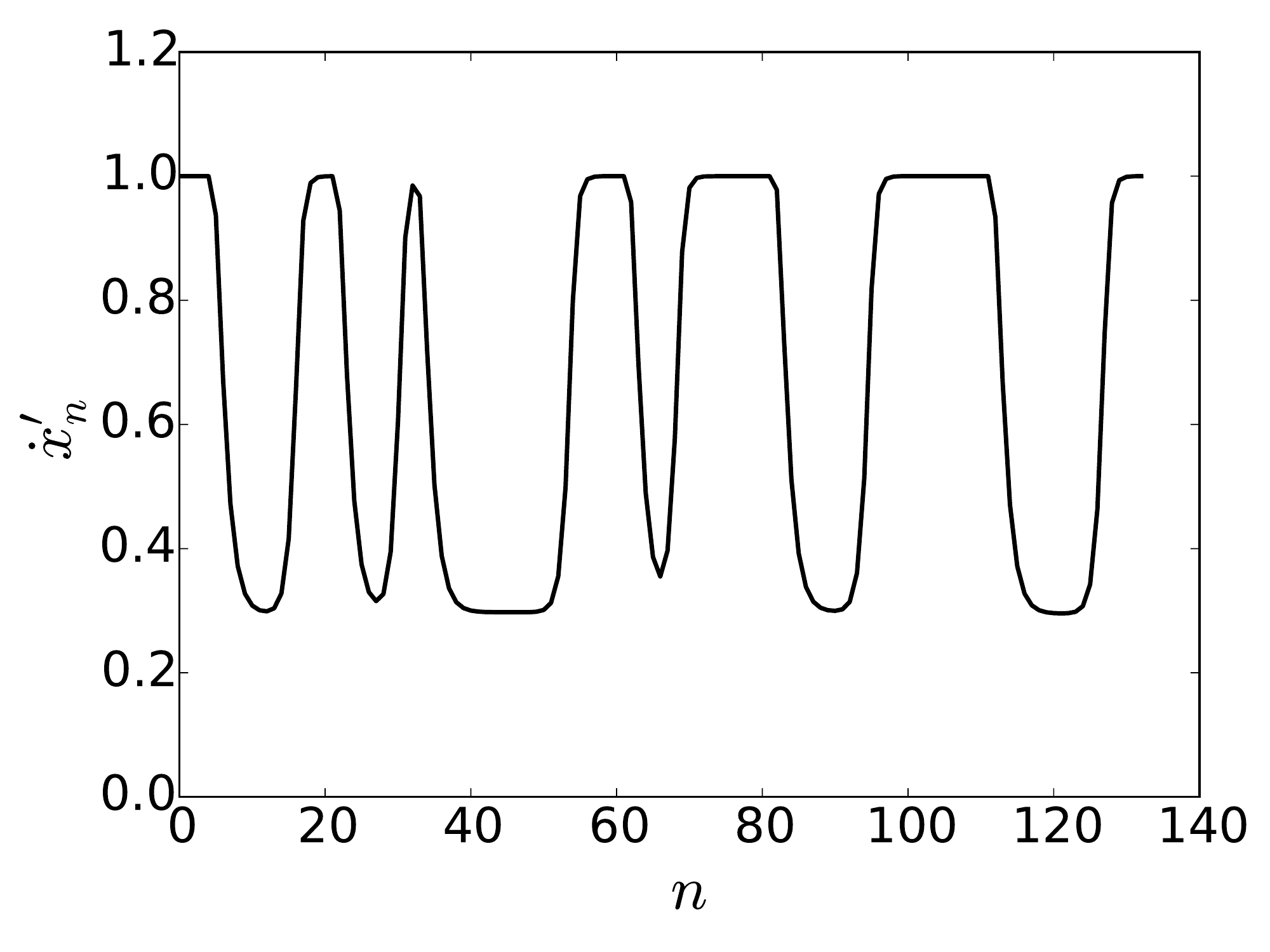}
\vspace{-0.4cm}
  \caption{Speed of pedestrians at different time steps. 
    Left: $t=300\,$ s,
    right: $t=2000\,$ s.}
  \label{fig:jams_vel}
\end{figure}

As shown in Fig.~\ref{fig:jams_vel} the speed does not become negative,
therefore backward movement is not observed.  This condition favors
the appearance of stable jams.

\begin{figure}[H]
  \centering
  \includegraphics[scale=0.32]{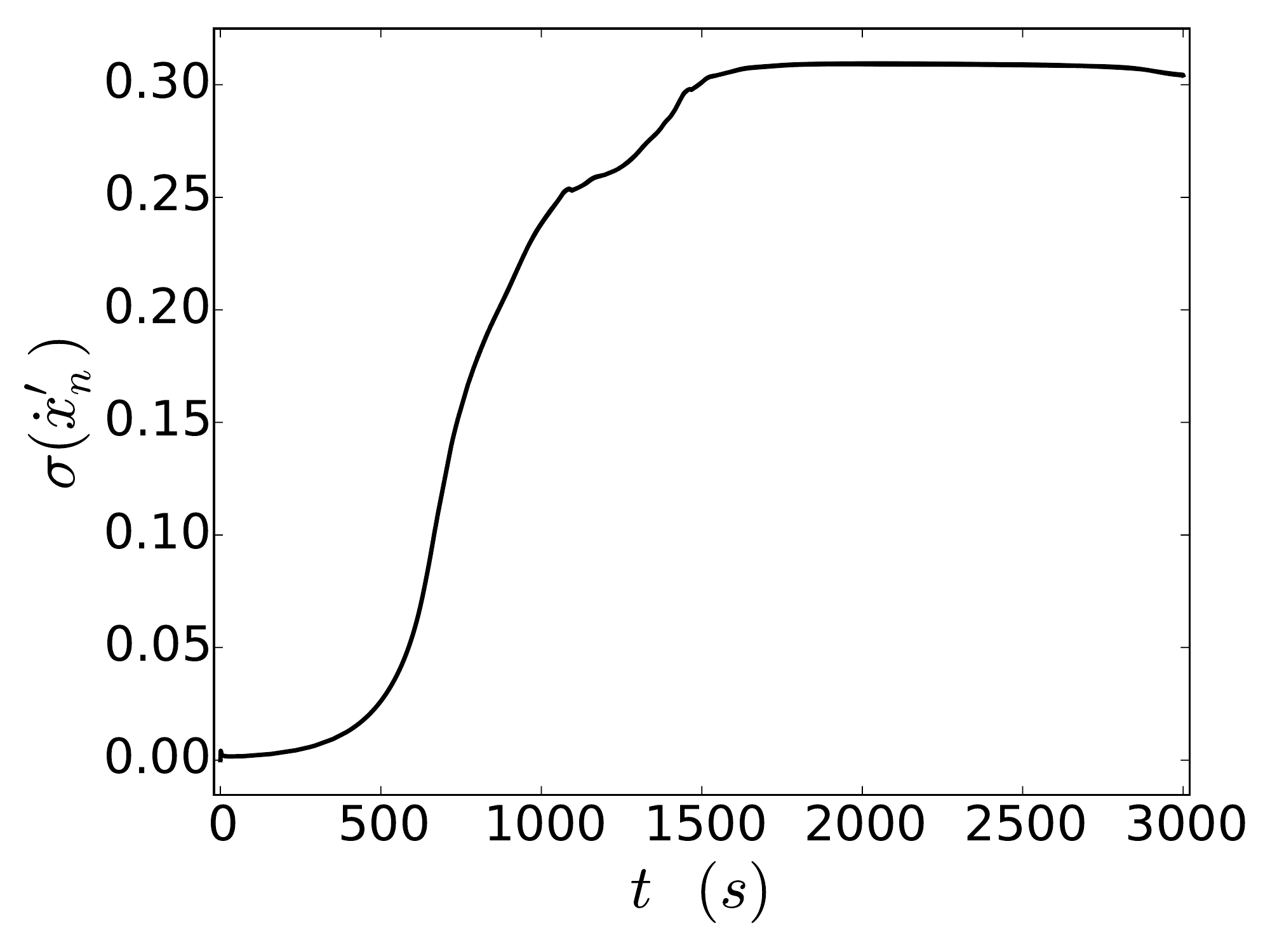}
  \caption{Standard deviation of the speed with respect to simulation time. The initial perturbation
      in the speed stabilizes at a non zero value.}
  \label{fig:std_vel_log}
\end{figure}

Fig.~\ref{fig:std_vel_log} shows the time evolution of the speed's standard deviation. After a relatively pronounced
 increase of the standard deviation, a stable plateau is formed. That means the system is in a ``stable'' homogeneous state.  

\section{Discussion and summary}
\label{sec:discussion}
Since their first application to pedestrian dynamics by Hirai and
Tarui~\cite{Hirai1977}, force-based models have been used extensively
to investigate the properties of crowds.  The ``goodness'' of these
models is usually asserted by means of qualitative and/or quantitative
investigations. Hereby, a model is judged to be realistic if its
description of pedestrian dynamics is consistent with empirical
findings. As example the fundamental diagram is often used as a
benchmark to test the plausibility of such models.

Depending on the expression of the repulsive force, we classify the
investigated force-based models as ``algebraically decaying'' and
``exponential-distance models''. The repulsive force in the first
category is inversely proportional to the effective distance of two
pedestrians
\cite{Yu2005,Chraibi2010a,Helbing2000a,Seyfried2006,Guo2010,Guo2012,Lohner2010,Shiwakoti2011}.
In the second category however, the magnitude of the repulsive force
increases exponentially with decreasing
distance~\cite{Helbing1995,Helbing2000,Lakoba2005,Johansson2007,Parisi2009,Moussaid2009}.
Hybrid models that rely on additional mechanisms to optimize the
desired direction of pedestrians (e.g.~\cite{Moussaid2011}) or to
handle collisions among pedestrians like for
example~\cite{Karamouzas2009,Karamouzas2014}, where the concept of the
time-to-collision is incorporated in the repulsive forces, make the
analytic form of the repulsive force way too complicated to be
investigated analytically.  Therefore, we do not include these models
in our analysis.

In this work we apply a method that gives new insights into the
characteristics of force-based models for pedestrian dynamics. It is
based on an analytical approach by investigating the linear stability 
of the homogeneous steady state. 
In this manner, it is possible to determine for which
parameter set, if any exists, a model is able to reproduce inhomogeneous states. 
Yet the nature of the unstable states (and the presence of realistic stop-and-go waves) 
has to be described by simulation.    
From an empirical point of view, the stop-and-go waves that were
observed in experiments under laboratory conditions
\cite{Portz2010,Lemercier2012} have a short pseudo-period. Hence, it
is not clear if these waves disappear after a long time or remain. In
all cases, their existence has been observed frequently in experiments
under laboratory conditions.

We have confirmed the analytical results by simulations which also
  give information about the nature of the unstable state.  These
  simulations have clearly shown that the unstable regions in the
investigated models do not show stop-and-go waves, but instead 
unrealistic behavior, e.g.\ backward movement and hence overlapping of
pedestrians. 

We have discussed that the superposition of forces may lead to
negative ``desired'' speeds and hence to backward movements.  In an
attempt to avoid this side-effect we have introduced a simple force-based
model that shows no negative speeds in simulations. As expected, the
model is able to produce stop-and-go waves in the instability region
instead. However, depending on the
chosen values for $v'_0$, collisions \textit{can} occur, as a result of backwards movement
and negative speeds.
 This is
explained by the fact that at the time $t_0$ when the sum of the
repulsive force and the positive driving term vanishes the system is
described by the following ODE
\begin{equation}
  \ddot x'_n + \dot x'_n = 0,
  \label{fig:decay_sys}
\end{equation}
which yields a speed that decays exponentially: 
\begin{equation}
  \dot x'_n = \dot x'_n(t_0)\exp(-t).
\end{equation}
$t_0$ can be interpreted as the time at which pedestrians start
anticipating possible collision.
Larger $v'_0$ implies a slower relaxation of the velocity.
Therefore, a possible enhancement of this model could be to shift the
minimal distance such that at $t_0$, $\Delta x'_n \ne 0$. That 
improves the ability of the system to tolerate 
slower decay of speeds for $t>t_0$. However, the 
main difficulty is that the value of the critical time $t_0$
remains unknown and can not be easily calculated.  This would 
require adding more complexity to the model, 
e.g. by considering behavioral anticipation of the dynamics, adding 
more (physical) forces or implementing extra collision detection 
techniques.

The investigations presented here were performed for single-file
motion, i.e. a strictly one-dimensional scenario. Although this situation
is well studied empirically in several controlled experiments, generically
pedestrian dynamics is two-dimensional. It remains to be seen, both
theoretically and empirically, how the scenario found here changes 
in this case.

\section{Appendix}
\subsection{Derivation of stability condition for algebraic forces}
\label{appA}

Here we give the details of the derivation of the stability criterion
of Sec.~\ref{sub-stab1}.

From (\ref{eq-pertub}) we find that
\begin{equation}
  \dot x_n' = v' + \dot \epsilon_n,\;\quad 
  \Delta \dot x_n' = \Delta \dot \epsilon_n\,,
  \quad 
  \ddot x_n' =  \ddot \epsilon_n,
  \label{eq-pertub2}
\end{equation}
since $\ddot y_n=0$. Inserting this into
the equation of motion Eq.~(\ref{eq:invd}) we obtain
\begin{equation}
  \ddot \epsilon_n =-F\cdot G+ v_0'-v' - \dot  \epsilon_n,
  \label{eqmotion}
\end{equation}
where $F$ and $G$ are defined as
\begin{eqnarray}
  F &=& \Big(d' + \Delta \epsilon_n  - \tilde a_v
        (\dot \epsilon_n + \dot \epsilon_{n+1}) \Big)^{-q}\\
  G &=& \Big({\mu + \delta r_\varepsilon(\Delta \dot \epsilon_n)}\Big)^2,
\end{eqnarray}
and $d' =\Delta y-2\tilde a_v v' - 2$.
We suppose that $v$ and $\rho$ are such that $d'\not=0$.

Considering the first-order approximation of $\exp(x)$ for 
$x\ll \varepsilon$  we have
\begin{equation}
  r_\varepsilon(x)  \approx \varepsilon \ln\Big(2-\frac{x}{\varepsilon}\Big)
  = \varepsilon \Big(\ln(2) + \ln( 1- \frac{x}{2\varepsilon} )\Big)
  \approx \varepsilon\ln(2) - \frac{1}{2}x\,.
  \label{eq:tanh}
\end{equation}
Then, 
\begin{equation}
  G \approx \left(\mu +  \delta\varepsilon\ln(2)
    - \frac{1}{2}\delta\Delta \dot \epsilon_n\right)^2
  \approx \gamma^2 - \delta\gamma  \Delta \dot \epsilon_n\,.
\end{equation}
where we have introduced $\gamma=\mu +  \delta\varepsilon\ln(2)$.
Using the effective distance Eq.~(\ref{eq:effDistn}), 
the expression for $F$ can be written as
\begin{equation}
  F= \Big({\frac{1}{d^\prime}\Big)^q\Big(1-\underbrace{\frac{\tilde a_v 
        (\dot \epsilon_n + \dot \epsilon_{n+1})-\Delta \epsilon_n}{d^\prime} }_{\ll 1}
    \Big)^{-q}}
  \approx \Big(\frac{1}{d^\prime}\Big)^q\Big(1+ q\frac{\tilde a_v (\dot 
    \epsilon_n + \dot \epsilon_{n+1})-\Delta \epsilon_n}{d^\prime}\Big).
\end{equation}
Substituting the expressions for $F$ and $G$ in Eq.~(\ref{eqmotion}) yields
\begin{equation}
  \ddot \epsilon_n = -\Big(\frac{1}{d^\prime}\Big)^q\Big( \gamma^2 +
  \frac{{\gamma}^2 q\tilde a_v}{{d^\prime}}(\dot \epsilon_n + \dot
  \epsilon_{n+1})  - \frac{{\gamma}^2q}{{d^\prime}}\Delta \epsilon_n
  -\delta\gamma\Delta \dot \epsilon_n \Big)
  + v_0' -v' -\dot \epsilon_n.
  \label{eq:x}
\end{equation}

In the steady state the equation of motion (\ref{eq:invd}) simplifies to
\begin{equation}
  0 = -\frac{\gamma^2}{{d^\prime}^q} + v_0'- v'\,,
  \label{eq2}
\end{equation}
and we obtain after rearranging Eq.~(\ref{eq:x})
\begin{equation}
  \ddot \epsilon_n = \frac{\delta\gamma}{{d^\prime}^q}\Delta \dot
  \epsilon_n + \frac{{\gamma}^2q}{{d^\prime}^{q+1}}\Delta \epsilon_n -
  \frac{{\gamma}^2q \tilde a_v}{{d^\prime}^{q+1}}(\dot \epsilon_n 
  + \dot \epsilon_{n+1})   - \dot \epsilon_n\,.
  \label{eq:epsf}
\end{equation}



Assuming a perturbation of the form
$\epsilon_n(t)=\alpha_ne^{z t}$ with $z\in\mathbb{C}$ and $\alpha_n\in\mathbb R$, $n=1,\ldots,N$, yields
\begin{equation}
  \alpha_nz^2 = \frac{\delta\gamma}{{d^\prime}^q}z(\alpha_{n+1}-\alpha_n) 
  + \frac{{\gamma}^2q}{{d^\prime}^{q+1}}(\alpha_{n+1}-\alpha_n) 
  - \frac{{\gamma}^2q \tilde a_v}{{d^\prime}^{q+1}}z(\alpha_n+\alpha_{n+1})  
  - \alpha_nz,
  \label{eq:a1}
\end{equation}
with $\alpha_{N+1}=\alpha_1$.
Introducing
\begin{equation}
  A=\frac{\delta\gamma}{{d^\prime}^q}z+\frac{{\gamma}^2q}{{d^\prime}^{q+1}}
  -\frac{{\gamma}^2q \tilde a_v}{{d^\prime}^{q+1}}z\qquad \mbox{and}\qquad 
  B=z^2+\frac{\delta\gamma}{{d^\prime}^q}z+\frac{{\gamma}^2q}{{d^\prime}^{q+1}}
  +\frac{{\gamma}^2q \tilde a_v}{{d^\prime}^{q+1}}z+z,
\end{equation}
Eq.~(\ref{eq:a1}) takes the simple form
\begin{equation}
  \alpha_n=\alpha_{n+1}\frac AB\,. 
\end{equation}
Iterating over $n$, we obtain the rational fraction in $z$
\begin{equation}
  \left(\frac AB\right)^N=1
  \qquad\Leftrightarrow\qquad 
  A=Be^{i2\pi l/N},
  \quad l=0,\ldots,N-1.
  \label{eq:a4}
\end{equation}
This equation is
\begin{equation}
  z^2 = \delta\gamma\frac{e^{\boldsymbol{i}k}-1}{{d^\prime}^q}z-
  \phi \tilde a_v(e^{\boldsymbol{i}k}+1)z
  +\phi(e^{\boldsymbol{i}k}-1)-z ,
  \label{eq:stab}
\end{equation}
with $\phi= \frac{q\gamma^2}{{d^\prime}^{q+1}}$ and 
$k=2\pi l/N$ with $l=0,\ldots,N-1$.

\begin{figure}[H]
  \begin{center}
    \subfigure{} 
    \includegraphics[width=0.4\columnwidth]{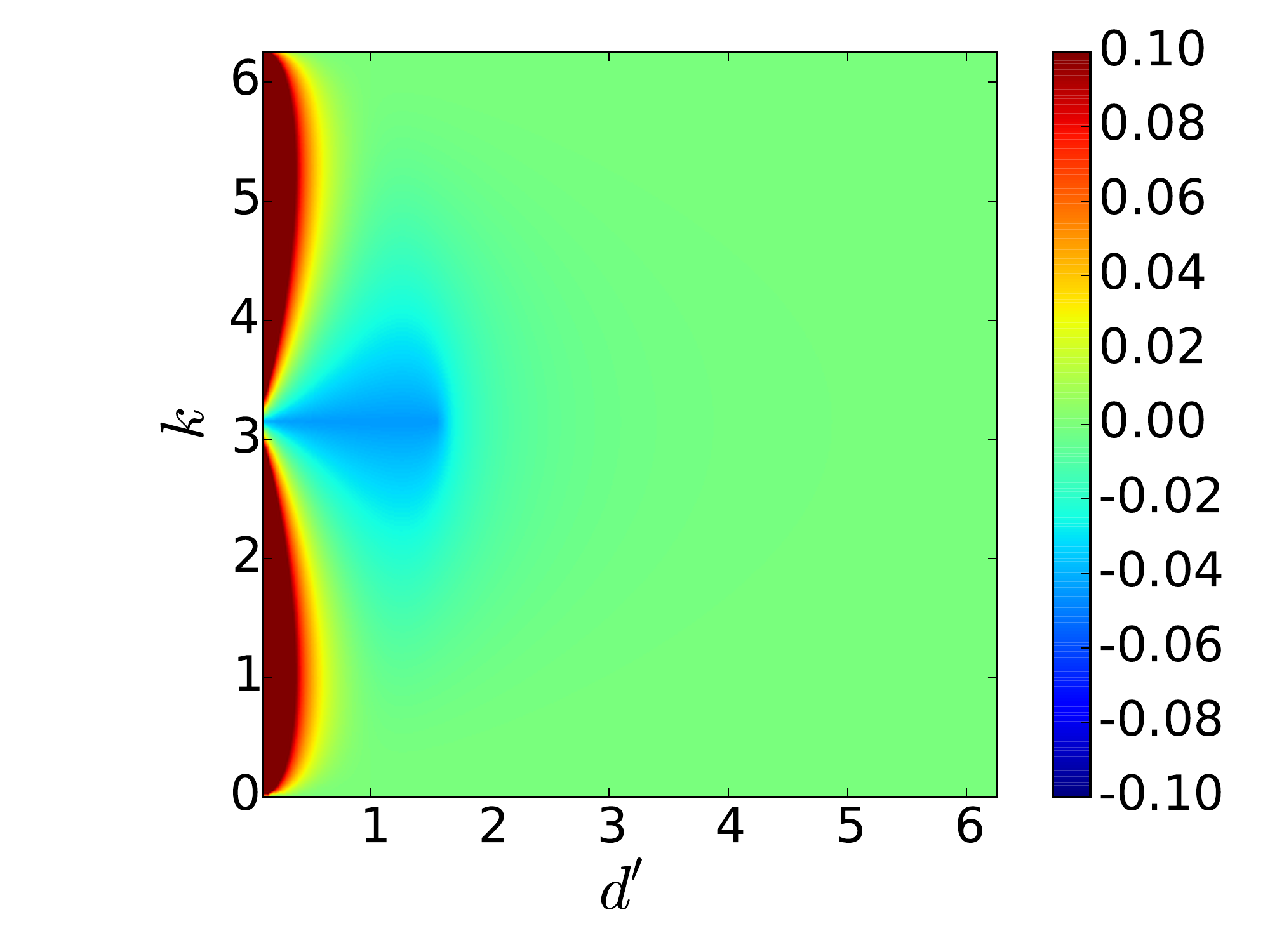}
    \subfigure{}
    \includegraphics[width=0.4\columnwidth]{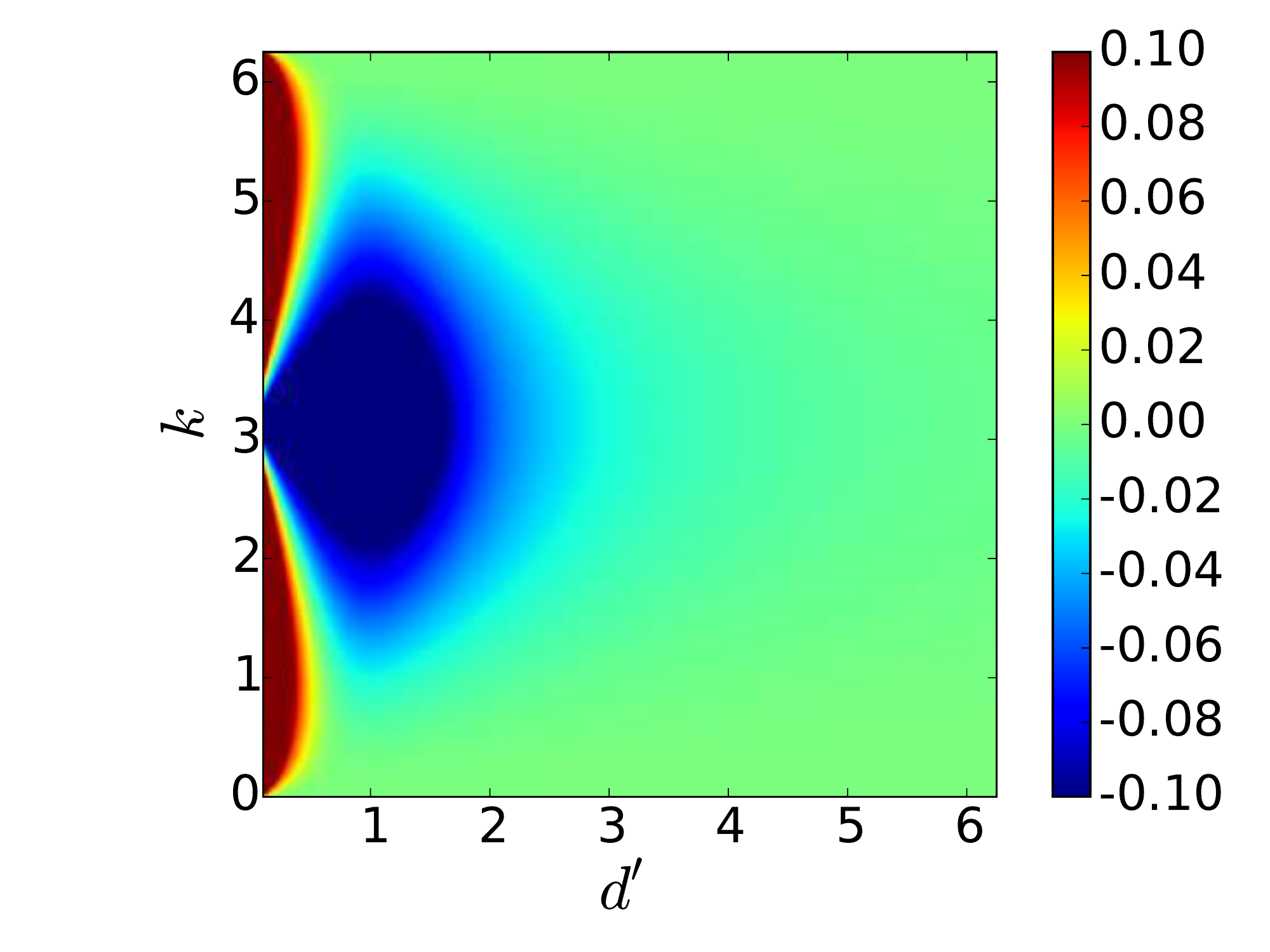}
    \subfigure{}
    \includegraphics[width=0.4\columnwidth]{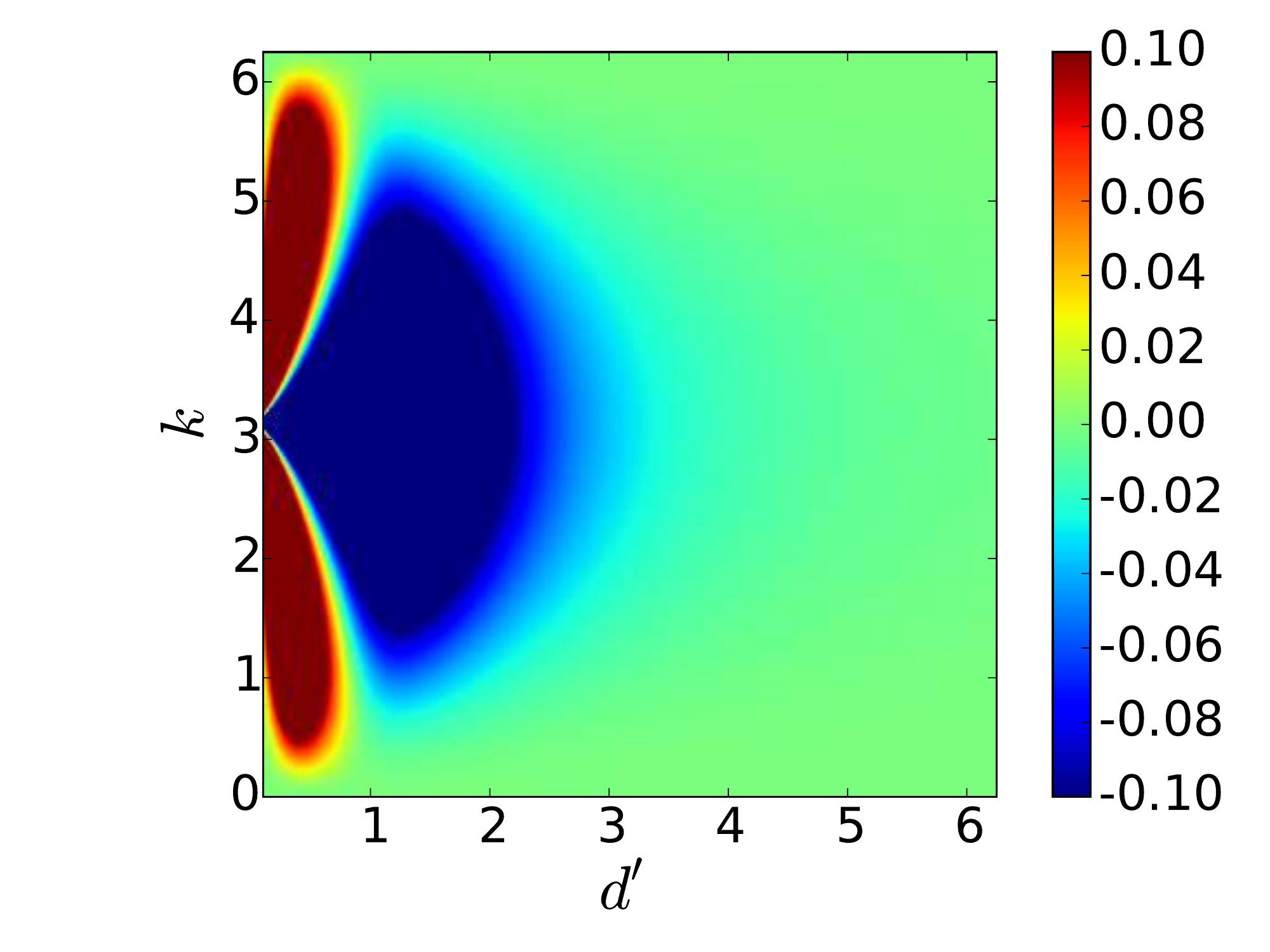}
    \subfigure{}
    \includegraphics[width=0.4\columnwidth]{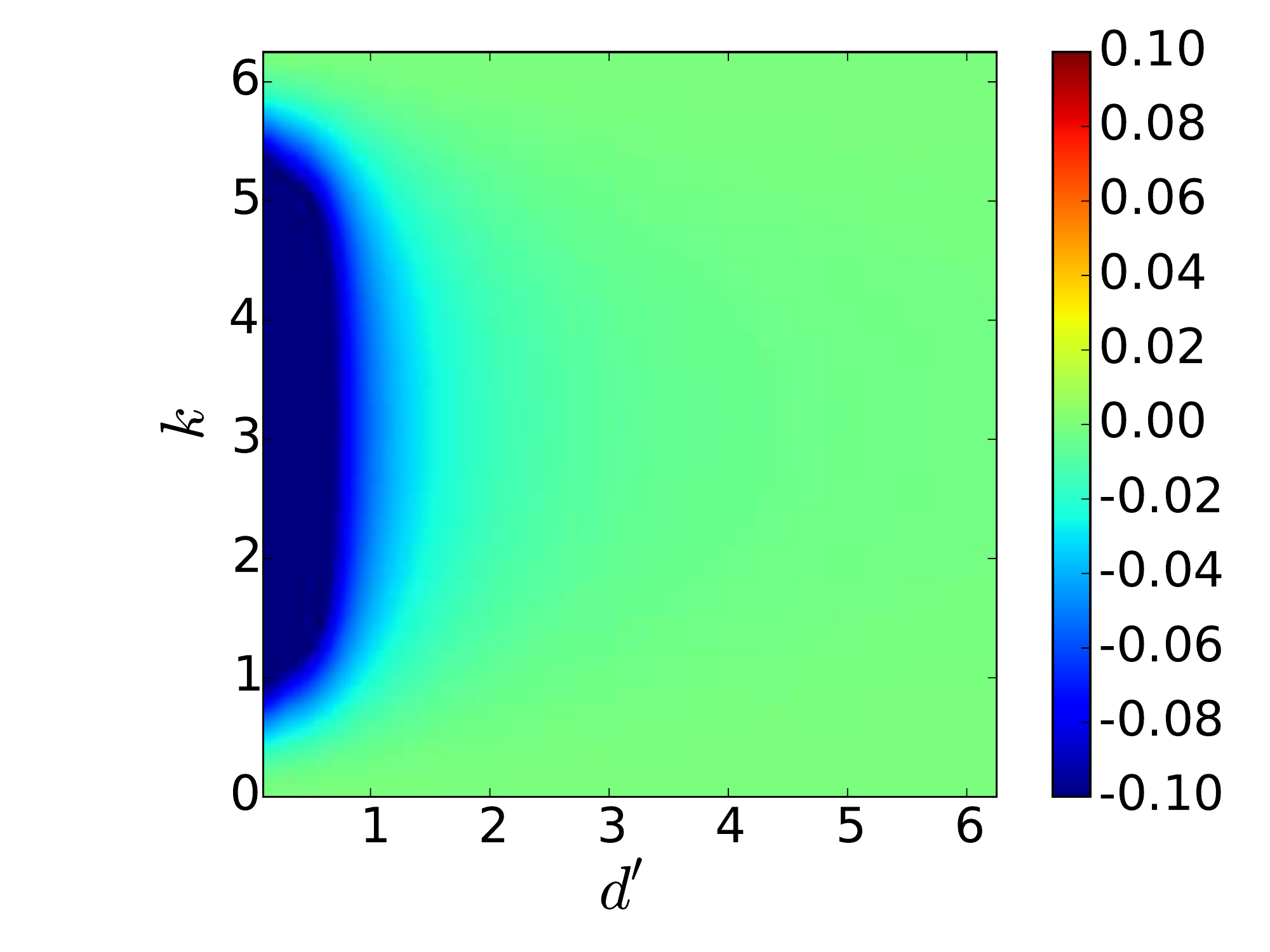}
    \caption{(Color online) Stability region in the $(d^\prime, k)$-space for
      different model classes. 
      Top left: $\mathcal{Q}= \langle 0.5, 0, 2,0\rangle$. 
      Top right: $\mathcal{Q}=  \langle 0.5, 0, 1, 0\rangle$.
      Bottom left: $\mathcal{Q}=  \langle 0.5, 0, 2, 0.1\rangle$.
      Bottom right: $\mathcal{Q}=  \langle 0.5, 1, 1, 0.1\rangle$. The colors
      are mapped to the value of $\Re [z^+]$ such that stability corresponds to  $\Re [z^+]<0$.
    }
    \label{fig:d-k}
  \end{center}
\end{figure}

The system described by the equation of motion ~(\ref{eq:invd}) is
stable if the real part $\Re [z]$ of all roots $z$ of
Eq.~(\ref{eq:stabN}) is negative.  Let $z^+$ and $z^-$ be two roots of
Eq.~(\ref{eq:stabN}). For five models  (see Tab.~I), we
investigate the stability regions in dependence of different wave
numbers $k$ and different densities (Fig.~\ref{fig:d-k}). Since $z^+>z^-$ it is enough to check the sign of $z^+$. 

We can observe that introducing a velocity-dependence in form of
relative velocity in the numerator of the repulsive term
(\ref{eq:invd}) or in the space requirement (\ref{eq:an}) has
a stabilizing effect on the behavior of the model, especially for
small wave numbers $k$.

\subsubsection{Stability for small $k$}
\label{appA2}

Limiting the expansion to second order and taking advantage of
$e^{\boldsymbol{i}k}\approx 1+\boldsymbol{i}k- \frac{k^2}{2}$ we obtain
from Eq.~(\ref{eq:stab})
\begin{align}
  {z^{(0)}}^2k^2 &= \frac{\delta\gamma}{{d^\prime}^q}\Big(
                   \boldsymbol{i}k-\frac{k^2}{2}\Big)\Big({z^{(0)}}k + {z^{(1)}}k^2\Big)
                   +\phi\Big(\boldsymbol{i}k-\frac{k^2}{2}\Big)\nonumber\\
                 &\quad - \tilde a_v\phi\Big(2+\boldsymbol{i}k-\frac{k^2}{2}\Big)
                   \Big({z^{(0)}}k + {z^{(1)}}k^2\Big)- ({z^{(0)}}k + {z^{(1)}}k^2)\nonumber\\
                 &=
                   \Big(\boldsymbol{i}\frac{\delta\gamma}{{d^\prime}^q}z^{(0)}-\frac{\phi}{2}
                   - 2 \tilde a_v\phi z^{(1)} -\boldsymbol{i} \tilde a_v\phi z^{(0)}
                   - z^{(1)}\Big)k^2
                   + \Big( \boldsymbol{i}\phi -2 \tilde a_v\phi z^{(0)} - z^{(0)}\Big)k.
\end{align}

Rearranging with respect to $k$ yields
\begin{equation}
  \Big({z^{(0)}}^2-\boldsymbol{i}\frac{\delta\gamma}{{d^\prime}^q}z^{(0)}
  + \frac{\phi}{2} + 2 \tilde a_v\phi{z^{(1)}} +\boldsymbol{i}\tilde a_v
  \phi z^{(0)}+ z^{(1)}\Big) k^2 -\Big( \boldsymbol{i}\phi
  -2\tilde a_v\phi z^{(0)} - z^{(0)}\Big)k=0.
  \label{eq:1}
\end{equation}

By a first-order approximation the terms with $k^2$ in
Eq.~(\ref{eq:1}) can be ignored which leads to
\begin{equation}
  \boldsymbol{i}\phi -2\tilde a_v\phi z^{(0)} - z^{(0)}=0.
  \label{eq:z00}
\end{equation}

Hence,
\begin{equation}
  z^{(0)} = \boldsymbol{i} \frac{\phi}{2\tilde a_v\phi + 1}.
  \label{eq:z0}
\end{equation}
With $\Re [z^{(0)}] =0$ we notice that a first order approximation is
not enough to provide the stability criterion, therefore we consider a
second order approximation. From Eq.~(\ref{eq:1}) and because of
Eq.~(\ref{eq:z00}) we obtain
\begin{equation}
  {z^{(0)}}^2-\boldsymbol{i}\Big(\frac{\delta\gamma}{{d^\prime}^q}-\tilde
  a_v\phi \Big) z^{(0)} + \Big( 2\tilde a_v\phi+1\Big)z^{(1)}+ \frac{\phi}{2} =0.
  \label{eq:2}
\end{equation}

Replacing the expression of $z^{(0)}$ from (\ref{eq:z0}) in (\ref{eq:2}) 
yields
\begin{equation}
  \Big(\boldsymbol{i} \frac{\phi}{2\tilde a_v\phi + 1}\Big)^2 -
  \boldsymbol{i}\Big(\frac{\delta\gamma}{{d^\prime}^q}-\tilde a_v\phi
  \Big)\Big(\boldsymbol{i} \frac{\phi}{2\tilde a_v\phi + 1}\Big)+
  \Big(2\tilde a_v\phi +1\Big)z^{(1)} + \frac{\phi}{2} =0,
\end{equation}
or
\begin{align}
  \frac{ 2\tilde a_v\phi+1}{\phi}z^{(1)} &= 
                                           \frac{\phi}{(2\tilde a_v\phi + 1)^2} -
                                           \Big(\frac{\delta\gamma}{{d^\prime}^q} -\tilde a_v\phi \Big)\Big(
                                           \frac{1}{2\tilde a_v\phi + 1}\Big)-\frac{1}{2},\;\; \phi \ne 0.
\end{align}
Since the coefficient of $z^{(1)}$ is positive, the system described
by Eq.~(\ref{eqmotion}) is linearly stable for $k\approx 0$ if
\begin{equation}
  \gamma>0,\qquad\phi\omega^2 -\Big(\frac{\delta\gamma}{{d^\prime}^q}- \tilde a_v \phi\Big)\omega - \frac{1}{2}<0, 
  \label{condition0}
\end{equation}
with the following notation $\omega=\frac{1}{2\tilde a_v\phi + 1}$.
Remarking that $\frac{1}{2}\omega(2\tilde a_v\phi + 1) = \frac{1}{2}$,
the inequality (\ref{condition0}) can be simplified to 
\begin{equation}
  \gamma>0,\qquad \Phi \coloneqq \phi\omega - \frac{\delta\gamma}{{d^\prime}^q} - \frac{1}{2}<0.
  \label{condition}
\end{equation}
Here, as a reminder, 
$\phi = \frac{q\gamma^2}{{d^\prime}^{q+1}}$,
$\gamma=\mu+\delta\varepsilon\ln(2)$ and $d'=\Delta y-2\tilde a_vv-2$,
with $\tilde a_v=a_v/\tau$. Note that since $\delta,\mu\ge0$ and
$\varepsilon>0$, $\gamma>0$ implies here $\mu>0$ or $\delta>0$.


\subsection{Derivation of stability condition for exponential forces}
\label{AppB}

As in the previous section we add a small dimensionless perturbation 
$\epsilon_n$ to the uniform solution and get from Eq.~(\ref{eq:sfm})
\begin{equation}
  \ddot \epsilon_n =-a \exp \Big(\frac{-d'}{b}\Big)
  \exp\Big(\frac{\tilde a_v(\dot \epsilon_n
    + \dot \epsilon_{n+1})-\Delta \epsilon_n}{b}\Big) - c\Big(\varepsilon
  \ln(2)-\frac{1}{2}(d'+\Delta  \epsilon - \tilde a_v(\dot \epsilon_n
  + \dot \epsilon_{n+1})\Big)+v_0' -   v' - \dot  \epsilon_n.
  \label{eq:seps}
\end{equation}
In the steady state we have $\ddot x^{(0)} = 0$ and
Eq.~(\ref{eq:sfm}) reduces to
\begin{equation}
  0 =  -a \exp\Big(\frac{-d'}{b}\Big) - c \Big(\varepsilon
  \ln(2)-\frac{1}{2}d'\Big) +v_0' -   v'\,.
  \label{eq:sfm_ss}
\end{equation}
Applying (\ref{eq:sfm_ss}) to (\ref{eq:seps}) yields
\begin{align}
  \ddot \epsilon_n&=\underbrace{-a \exp \Big(\frac{-d'}{b}\Big)}_{\tilde a} 
                    \Big(\exp\Big(\frac{\tilde a_v(\dot \epsilon_n
                    + \dot \epsilon_{n+1}) - \Delta \epsilon_n}{b}\Big) -1\Big) 
                    +\frac{1}{2}c\Big(\Delta\epsilon_n-\tilde a_v(\dot \epsilon_n
                    + \dot \epsilon_{n+1})\Big)  -\dot   \epsilon_n\nonumber\\
                  &\approx \tilde a\Big(\frac{\tilde a_v(\dot \epsilon_n
                    + \dot \epsilon_{n+1})-\Delta \epsilon_n}{b} \Big) +\frac{1}{2}c\Big(\Delta
                    \epsilon_n-\tilde a_v(\dot \epsilon_n
                    + \dot \epsilon_{n+1})\Big) -\dot \epsilon_n\nonumber\\
                  &=\tilde a_v(\tilde a/b-\frac{1}{2}c)(\dot \epsilon_n + \dot
                    \epsilon_{n+1})-(\tilde a/b - \frac{1}{2}c)\Delta \epsilon_n - \dot \epsilon_n.
\end{align}
By introducing the substitutions $\tilde c =\tilde a/b - \frac{1}{2}c$
and $\tilde b = \tilde a_v\tilde c$ we obtain a simplified equation for
the perturbation:
\begin{equation}
  \ddot \epsilon_n = \tilde b(\dot \epsilon_n + \dot
  \epsilon_{n+1}) - \tilde c\Delta \epsilon_n - \dot \epsilon_n.
\end{equation}
Using the expansion $\epsilon_n(t)=\alpha_n e^{zt}$, we obtain 
\begin{equation}
  z^2-\Big(\tilde b(e^{\boldsymbol{i}k}+1)  -1\Big)z 
  +\tilde c(e^{\boldsymbol{i}k}-1)=0.
  \label{eq:polyExp}
\end{equation}

\begin{figure}[H]
  \begin{center}
    \subfigure{}
     \includegraphics[width=0.4\columnwidth]{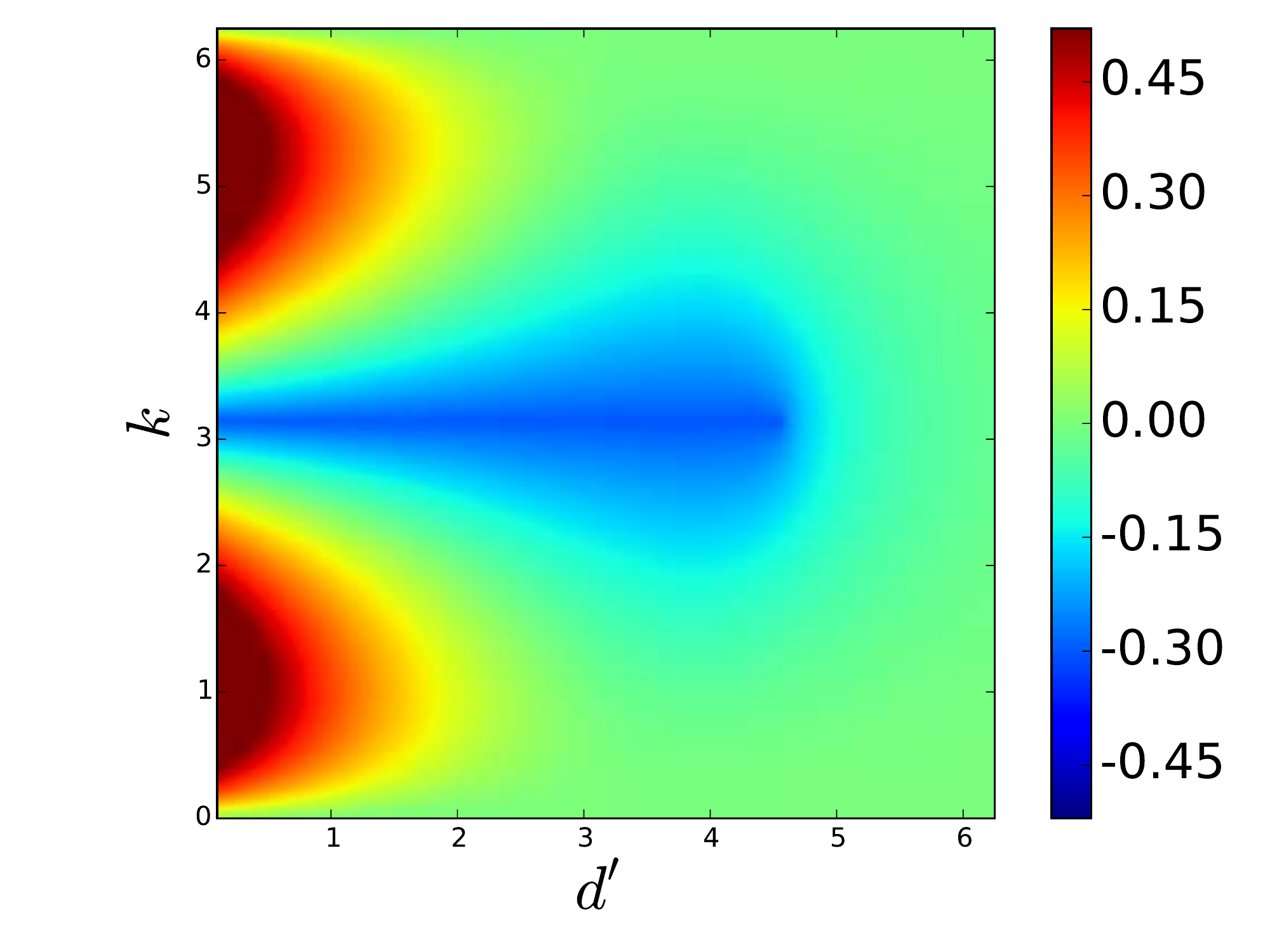}
    \subfigure{}
     \includegraphics[width=0.4\columnwidth]{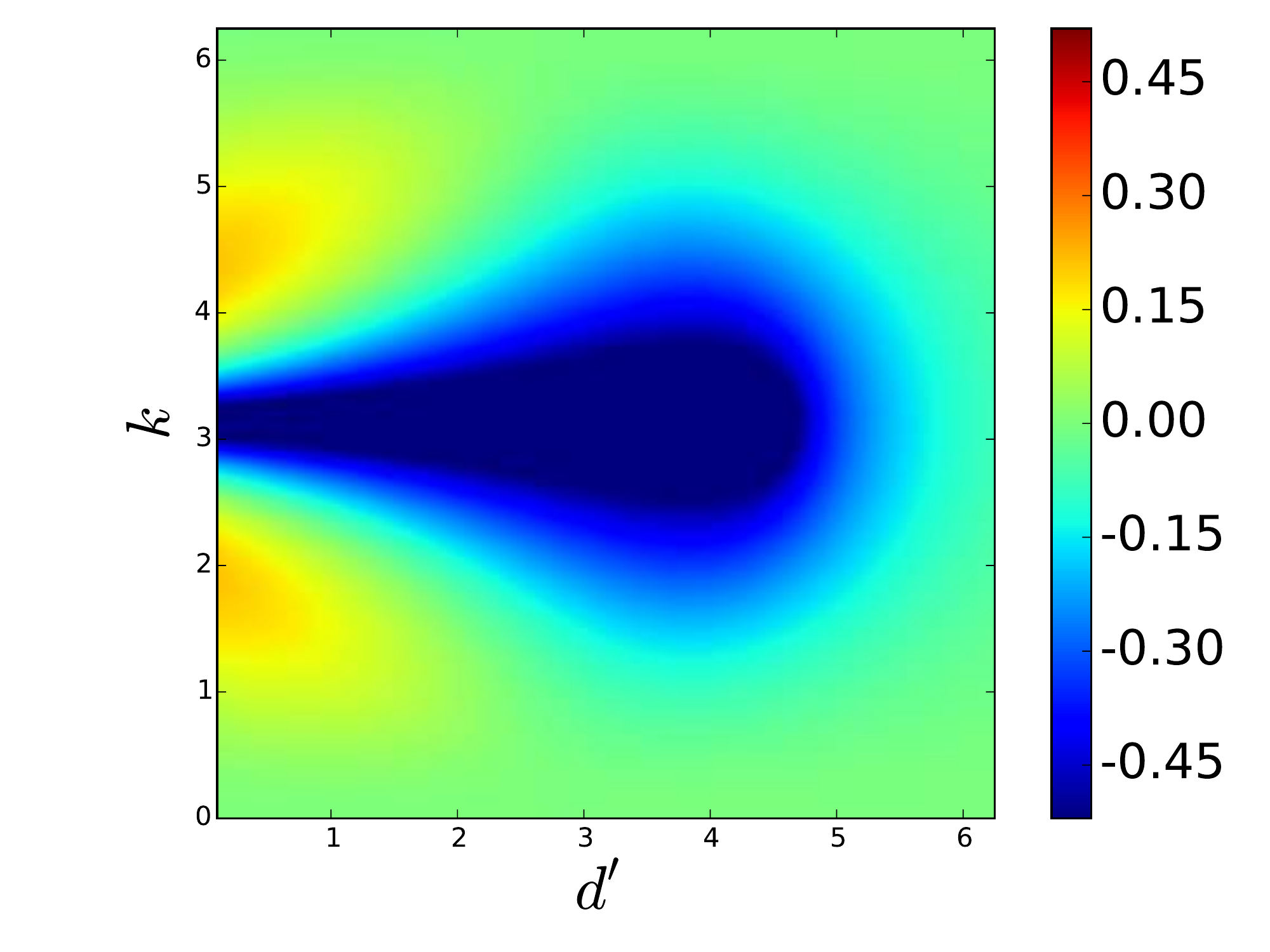}
     \includegraphics[width=0.4\columnwidth]{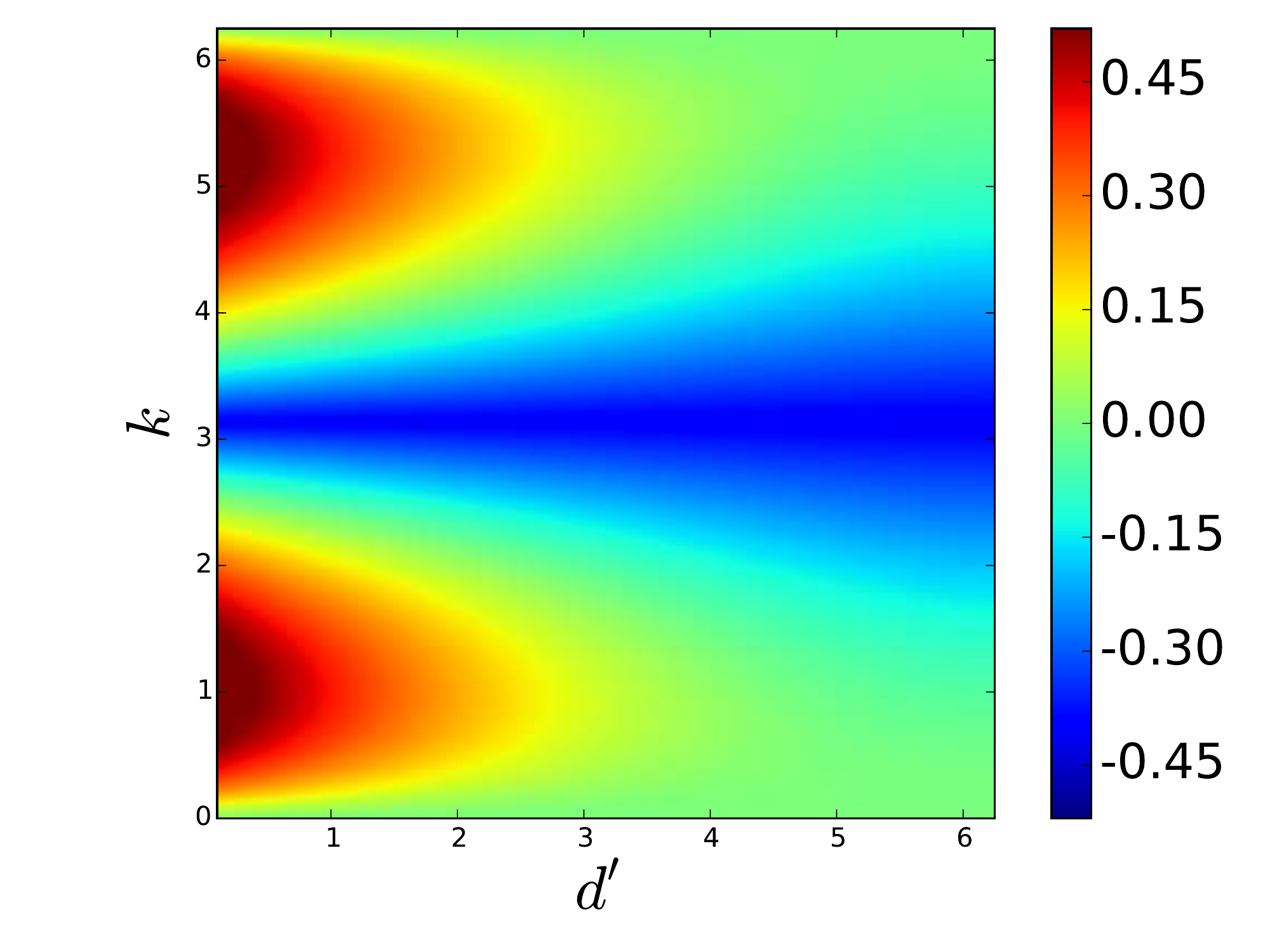}
    \subfigure{}
     \includegraphics[width=0.4\columnwidth]{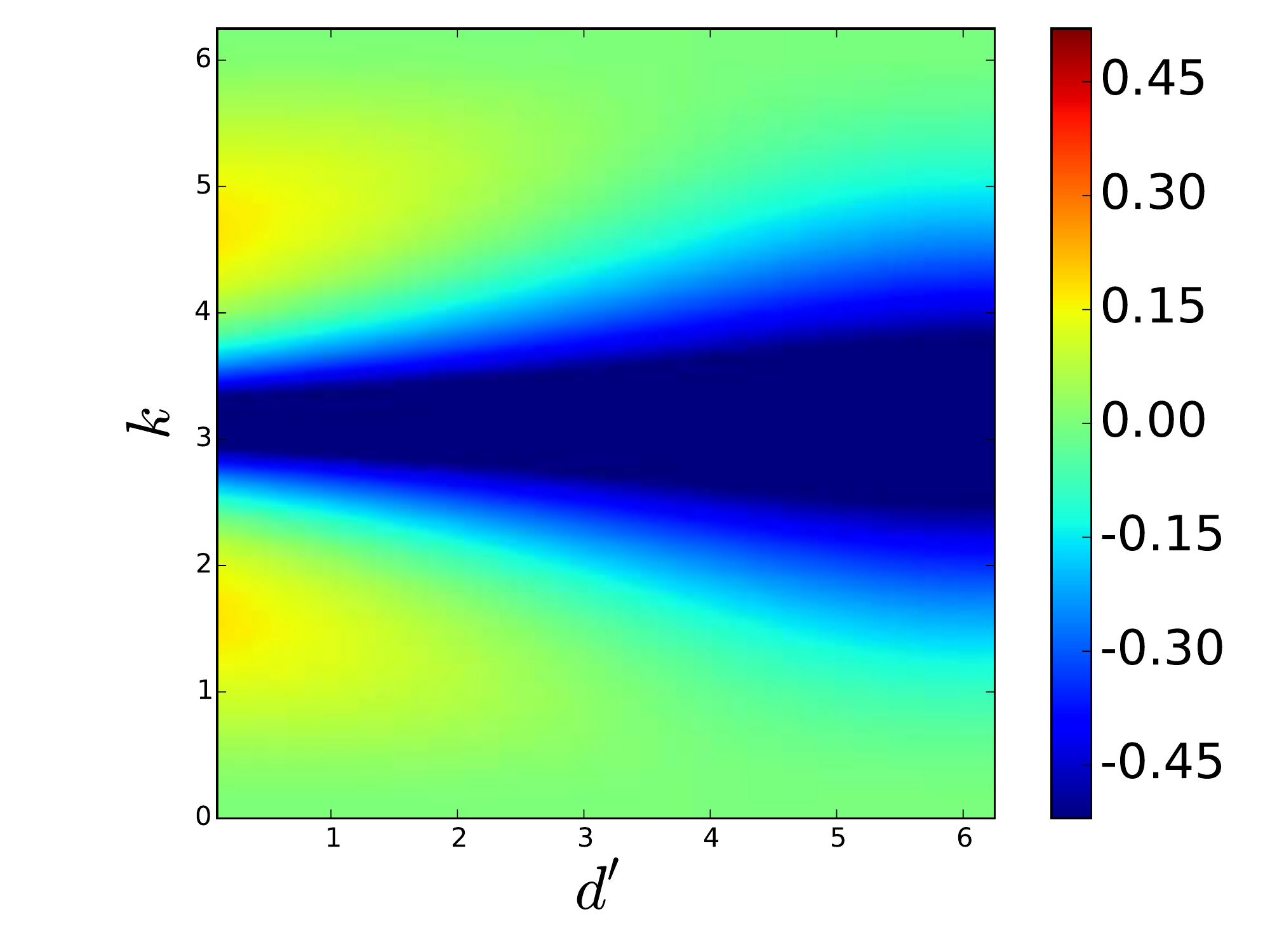}
    \caption{(Color online) Stability region in the $(d^\prime, k)$-space for
      different model classes.  Up left: $\mathcal{\tilde Q}= \langle
      12, 1, 0, 0\rangle$.  Up right: $\mathcal{\tilde Q}= \langle 12,
      1, 0, 0.2\rangle$.  Bottom left: $\mathcal{\tilde Q}= \langle 12,
      2, 0, 0\rangle$.  Bottom right: $\mathcal{\tilde Q}= \langle 12,
      2, 0, 0.2\rangle$.  The colors are mapped to the value of
      $\Re{(\tilde z^+)}$.
    }
    \label{fig:d-k-exp}
  \end{center}
\end{figure}

Fig.~\ref{fig:d-k-exp} shows the instability regions in the ($k, d'$)-space. 
With $\tilde a_v \ne 0$ the instability of the system is
considerably reduced. 
\subsubsection{Stability for small $k$}
\label{AppB2}

We further focus on the case  $k\approx 0$. For the solution
$z\approx z^{(0)}k + z^{(1)}k^2$ we obtain by substituting in (\ref{eq:polyExp})
\begin{align}
  {z^{(0)}}^2&=\tilde b(2 + \boldsymbol{i}k-\frac{k^2}{2})(z^{(0)}k 
               + z^{(1)}k^2) -\tilde
               c(\boldsymbol{i}k-\frac{k^2}{2})  -(z^{(0)}k +z^{(1)}k^2).
\end{align}
Rearranging the coefficients of the same power yields
\begin{equation}
  \Big(-{z^{(0)}}^2 + 2\tilde b z^{(1)} + \boldsymbol{i}\tilde b z^{(0)} 
  + \frac{\tilde c}{2} -
  z^{(1)} \Big)k^2 + \Big(2\tilde b z^{(0)} -\boldsymbol{i}\tilde c-z^{(0)}  
  \Big)k=0.
  \label{eq:pol}
\end{equation}
A first-order approximation yields by ignoring the $k^2$-term in 
(\ref{eq:pol}):
\begin{equation}
  z^{(0)} = \boldsymbol{i}\frac{\tilde c}{2\tilde b\tau -1}.
  \label{eq:z02}
\end{equation}
Since $\Re(z^{(0)})=0$, we consider a second order approximation of $z$. 
Therefore, replacing $z^{(0)}$ by its expression from (\ref{eq:z02}) yields
\begin{equation}
  z^{(1)}\left(2\tilde b-1\right)= -\frac{\tilde c}{2} -
  \left(\frac{\tilde c}{2\tilde b -1}\right)^2  +\tilde b 
  \frac{\tilde c}{2\tilde b -1}.
\end{equation}
Finally we obtain for $z^{(1)}$
\begin{equation}
  z^{(1)}= -\left(\frac{\tilde c}{2} + \left(\frac{\tilde c}{2\tilde
        b -1}\right)^2 - \tilde b \frac{\tilde c}{2\tilde b
      -1}\right)\Big(\frac{1}{2\tilde b-1}\Big)
  = -\alpha\left(\frac{\tilde c}{2} + \tilde c^2\alpha^2 
    - \tilde b \tilde c \alpha\right).
\end{equation}
and the system is linearly stable for $k\approx 0$ if
\begin{equation}
  -\alpha\left(\frac{\tilde c}{2} + \tilde c^2\alpha^2 
    - \tilde b \tilde c\alpha\right)<0
\end{equation}
where $\alpha=\frac{1}{2\tilde b-1}$.
By simplifying using $-\alpha^2\tilde c >0$, we obtain the condition
\begin{equation}
  \tilde \Phi=-\frac12+\tilde c \alpha<0,
\end{equation}
with $\alpha=\frac1{2\tilde b -1}$, $\tilde b =\tilde a_v \tilde c$, 
$\tilde c=\tilde a/b-\frac{1}{2}c$ and $\tilde a=-a\exp(-d'/b)$.

\begin{acknowledgments}
  M.C. is grateful to Japan Society for the Promotion of Science (JSPS)
  for funding this work under Grant-Nr.: PE 12078.  T.E. acknowledges
  support from JSPS Grants-in-Aid for Scientific Research (13J05086).
  A.Sch. thanks the Deutsche Forschungsgemeinschaft (DFG) for support
  under grant ``Scha 636/9-1''.
\end{acknowledgments}



\begin{thebibliography}{34}
\bibitem{Helbing2001}
D.~Helbing.
\newblock {Traffic and related self-driven many-particle systems}.
\newblock {\em Rev. Mod. Phys.}, 73:1067--1141, 2001.

\bibitem{Schadschneider2009a}
A.~Schadschneider, W.~Klingsch, H.~Kl\"upfel, T.~Kretz, C.~Rogsch, and
  A.~Seyfried.
\newblock {\em {Encyclopedia of Complexity and System Science}}, volume~5,
  pages 3142--3176.
\newblock Springer, Berlin Heidelberg, 2009.

\bibitem{Schadschneider2010b}
A.~Schadschneider, D.~Chowdhury, and K.~Nishinari.
\newblock {\em Stochastic Transport in Complex Systems. From Molecules to
  Vehicles}.
\newblock Elsevier Science Publishing Co Inc., 2010.

\bibitem{Ali2013}
Saad Ali, Ko~Nishino, Dinesh Manocha, and Mubarak Shah, editors.
\newblock {\em Modeling, Simulation and Visual Analysis of Crowds A
  Multidisciplinary Perspective}.
\newblock Springer New York, 2013.

\bibitem{Seyfried2008}
A.~Seyfried and A.~Schadschneider.
\newblock {Fundamental Diagram and Validation of Crowd Models}.
\newblock {\em Lect. Notes Comp. Sc.}, 5191:563--566, 2008.

\bibitem{Schadschneider2009c}
Andreas Schadschneider and Armin Seyfried.
\newblock {Empirical Results for Pedestrian Dynamics and their Implications for
  Cellular Automata Models}.
\newblock In Harry Timmermans, editor, {\em {Pedestrian Behavior: Data
  Collection and Applications}}, chapter~2, pages 27--43. Emerald Group
  Publishing Limited, 1 edition, nov 2009.

\bibitem{ZhangQ2011}
Qi~Zhang and Baoming Han.
\newblock Simulation model of pedestrian interactive behavior.
\newblock {\em Physica A}, 390:636--646, 2011.

\bibitem{Yu2005}
W.~J. Yu, R.~Dong Chen, L.Y., and S.Q. Dai.
\newblock {Centrifugal force model for pedestrian dynamics}.
\newblock {\em Phys. Rev. E}, 72(2):026112, 2005.

\bibitem{Helbing2004}
D.~Helbing.
\newblock {Collective phenomena and states in traffic and self-driven
  many-particle systems}.
\newblock {\em Comp. Mater. Sci.}, 30(1--2):180--187, 2004.

\bibitem{Lakoba2005}
T.~I. Lakoba, D.~J. Kaup, and N.~M. Finkelstein.
\newblock {Modifications of the Helbing-Moln\'{a}r-Farkas-Vicsek social force
  model for pedestrian evolution}.
\newblock {\em Simulation}, 81(5):339--352, 2005.

\bibitem{Parisi2007}
D.~R. Parisi and C.~O. Dorso.
\newblock {Morphological and dynamical aspects of the room evacuation process}.
\newblock {\em Physica A}, 385(1):343--355, 2007.

\bibitem{Garcimartin2014}
A.~Garcimartin, I.~Zuriguel, J.~M. Pastor, C.~Mart\i~n G\o~mez, and D.~R.
  Parisi.
\newblock Experimental evidence of the "faster is slower" effect.
\newblock {\em Transportation Research Procedia}, 2(0):760 -- 767, 2014.

\bibitem{Parisi2015}
D.~R. Parisi, S.~A. Soria, and R.~Josens.
\newblock Faster-is-slower effect in escaping ants revisited: Ants do not
  behave like humans.
\newblock {\em Safety Science}, 72(0):274 -- 282, 2015.

\bibitem{Portz2010}
Andrea Portz and Armin Seyfried.
\newblock {Modeling stop-and-go waves in pedestrian dynamics}.
\newblock In Roman Wyrzykowski, Jack Dongarra, Konrad Karczewski, and Jerzy
  Wasniewski, editors, {\em {PPAM 2009, Part II}}, pages 561--568, Berlin
  Heidelberg, 2010. Springer.

\bibitem{Seyfried2010b}
Armin Seyfried, Andrea Portz, and Andreas Schadschneider.
\newblock {Phase coexistence in congested sates of pedestrian dynamics}.
\newblock {\em Lect. Notes Comp. Sc.}, 6350:496--505, 2010.

\bibitem{Lemercier2012}
Samuel Lemercier, Asja Jelic, Richard Kulpa, Jiale Hua, J\'er\^ome Fehrenbach,
  Pierre Degond, C\'ecile Appert-Rolland, St\'ephane Donikian, and Julien
  Pettr\'e.
\newblock Realistic following behaviors for crowd simulation.
\newblock {\em Computer Graphics Forum}, 31:489--498, 2012.

\bibitem{Eilhardt2014}
Christian Eilhardt and Andreas Schadschneider.
\newblock Stochastic headway dependent velocity model for 1d pedestrian
  dynamics at high densities.
\newblock {\em Transportation Research Procedia}, 2(0):400 -- 405, 2014.

\bibitem{Chraibi2014a}
Mohcine Chraibi.
\newblock Oscillating behavior within the social force model.
\newblock {\em e-print arXiv:1412.1133}, 2014.

\bibitem{Chowdhury2000}
Debashish Chowdhury, Ludger Santen, and Andreas Schadschneider.
\newblock {Statistical physics of vehicular traffic and some related systems}.
\newblock {\em Phys. Rep.}, 329(4--6):199--329, 2000.

\bibitem{Gazis2002}
D.~C. Gazis.
\newblock The origins of traffic theory.
\newblock {\em Op Res.}, 50(1):69--77, 2002.

\bibitem{Orosz2010}
G.~Orosz, R.~E. Wilson, and G.~Stepan.
\newblock Traffic jams~: dynamics and control.
\newblock {\em Proc. R. Soc. A}, 368(1957):4455--4479, 2010.

\bibitem{Nagatani2002a}
T.~Nagatani.
\newblock The physics of traffic jams.
\newblock {\em Rep. Prog. Phys.}, 65(9):13--31, 2002.

\bibitem{Koester2013}
Gerta K\"oster, Franz Treml, and Marion G\"odel.
\newblock Avoiding numerical pitfalls in social force models.
\newblock {\em Phys. Rev. E}, 87, 2013.

\bibitem{Chraibi2014}
Mohcine Chraibi, Armin Seyfried, and Andreas Schadschneider.
\newblock Quantitative validation of the generalized centrifugal force model.
\newblock In Ulrich Weidmann, Uwe Kirsch, and Michael Schreckenberg, editors,
  {\em Pedestrian and Evacuation Dynamics 2012}, pages 603--613. Springer,
  2014.

\bibitem{Chraibi2010a}
Mohcine Chraibi, Armin Seyfried, and Andreas Schadschneider.
\newblock {The generalized centrifugal force model for pedestrian dynamics}.
\newblock {\em Phys. Rev. E}, 82:046111, 2010.

\bibitem{Chraibi2011}
Mohcine Chraibi, Ulrich Kemloh, Armin Seyfried, and Andreas Schadschneider.
\newblock Force-based models of pedestrian dynamics.
\newblock {\em Networks and Heterogeneous Media}, 6(3):425--442, 2011.

\bibitem{Berg2008}
Jur van~den Berg, Ming Lin, and Dinesh Manocha.
\newblock Reciprocal velocity obstacles for real-time multi-agent navigation.
\newblock In {\em {IEEE International Conference on Robotics and Automation,
  2008. ICRA 2008}}, pages 1928--1935, 2008.

\bibitem{Maury2009}
B.~Maury and J.~Venel.
\newblock {Handling of contacts on crowd motion simulations}.
\newblock In {\em Traffic and Granular Flow '07}. Springer, 2009.

\bibitem{Venel2010}
J.~Venel.
\newblock Integrating strategies in numerical modelling of crowd motion.
\newblock In {\em Pedestrian and Evacuation Dynamics 2008}, 2010.

\bibitem{Patil2010}
Sachin Patil, Jur van~den Berg, Sean Curtis, Ming Lin, and Dinesh Manocha.
\newblock {Directing Crowd Simulations Using Navigation Fields}.
\newblock {\em IEEE Transactions On Visualization And Computer Graphics}, 16,
  2010.

\bibitem{Dietrich2014}
Felix Dietrich and Gerta K\"oster.
\newblock Gradient navigation model for pedestrian dynamics.
\newblock {\em Phys. Rev. E}, 89:062801, Jun 2014.

\bibitem{Dietrich2014a}
Felix Dietrich, Gerta K{\"o}ster, Michael Seitz, and Isabella von Sivers.
\newblock Bridging the gap: From cellular automata to differential equation
  models for pedestrian dynamics.
\newblock {\em J. Comp. Sc.}, 5(5):841 -- 846, 2014.

\bibitem{Kirik2014}
Ekaterina Kirik and Andrey Malyshev.
\newblock On validation of sigmaeva pedestrian evacuation computer simulation
  module with bottleneck flow.
\newblock {\em Journal of Computational Science}, 5:847--850, 2014.

\bibitem{Portz2011}
Andrea Portz and Armin Seyfried.
\newblock Analyzing stop-and-go waves by experiment and modeling.
\newblock In R.D. Peacock, E.D. Kuligowski, and J.D. Averill, editors, {\em
  Pedestrian and Evacuation Dynamics 2010}, pages 577--586. Springer, 2011.

\bibitem{Weidmann1993}
U.~Weidmann.
\newblock {Transporttechnik der Fussg\"anger}.
\newblock Technical Report Schriftenreihe des IVT Nr. 90, Institut f\"ur
  Verkehrsplanung,Transporttechnik, Strassen- und Eisenbahnbau, ETH Z\"urich,
  ETH Z\"urich, 1993.
\newblock 2nd Edition.

\bibitem{Chraibi2012a}
Mohcine Chraibi, Martina Freialdenhoven, Andreas Schadschneider, and Armin
  Seyfried.
\newblock Modeling the desired direction in a force-based model for pedestrian
  dynamics.
\newblock In {\em Traffic and Granular Flow'11}, pages 263--275. Springer
  Berlin Heidelberg, 2013.

\bibitem{Note1}
In GCFM pedestrians are modeled by ellipses with two velocity-dependent
  semi-axes.

\bibitem{Helbing2000a}
D.~Helbing, I.~J. Farkas, and T.~Vicsek.
\newblock {Freezing by heating in a driven mesoscopic system}.
\newblock {\em Phys. Rev. Lett.}, 84:1240--1243, 2000.

\bibitem{Seyfried2006}
A.~Seyfried, B.~Steffen, and T.~Lippert.
\newblock {Basics of modelling the pedestrian flow}.
\newblock {\em Physica A}, 368:232--238, 2006.

\bibitem{Guo2010}
Ren-Yong Guo, S.~C. Wong, Hai-Jun Huang, Zhang Peng, and William H.~K. Lam.
\newblock {A microscopic pedestrian-simulation model and its application to
  intersecting flows}.
\newblock {\em Physica A}, 389(3):515--526, feb 2010.

\bibitem{Guo2012}
Ren-Yong Guo and Tie-Qiao Tang.
\newblock A simulation model for pedestrian flow through walkways with corners.
\newblock {\em Simulation Modelling Practice and Theory}, 21:103 -- 113, 2012.

\bibitem{Lohner2010}
Rainald L\"ohner.
\newblock {On the modelling of pedestrian motion}.
\newblock {\em Appl. Math. Model.}, 34(2):366--382, 2010.

\bibitem{Shiwakoti2011}
N.~Shiwakoti, M.~Sarvi, G.~Rose, and M.~Burd.
\newblock Animal dynamics based approach for modelling pedestrian crowd egress
  under panic conditions.
\newblock {\em Transportation and Traffic Theory}, 17:438--461, 2011.

\bibitem{Karamouzas2014}
Ioannis Karamouzas, Brian Skinner, and Stephen~J. Guy.
\newblock A universal power law governing pedestrian interactions.
\newblock {\em Phys. Rev. Lett.}, 113(5):238701, Dec 2014.

\bibitem{Moussaid2011}
Mehdi Moussa\"id, Dirk Helbing, and Guy Theraulaz.
\newblock How simple rules determine pedestrian behavior and crowd disasters.
\newblock {\em P. Natl. Acad. Sci. USA.}, 108(17):6884--6888, 2011.

\bibitem{Treiber2015}
Martin Treiber and Venkatesan Kanagaraj.
\newblock Comparing numerical integration schemes for time-continuous
  car-following models.
\newblock {\em Physica A: Statistical Mechanics and its Applications},
  419(0):183--195, 2015.

\bibitem{Helbing1995}
D.~Helbing and P.~Moln\'{a}r.
\newblock {Social force model for pedestrian dynamics}.
\newblock {\em Phys. Rev. E}, 51:4282--4286, 1995.

\bibitem{Helbing2000}
D.~Helbing, I.~Farkas, and T.~Vicsek.
\newblock {Simulating dynamical features of escape panic}.
\newblock {\em Nature}, 407:487--490, 2000.

\bibitem{Johansson2007}
Anders Johansson, Dirk Helbing, and Pradyumn~K. Shukla.
\newblock {Specification of the social force pedestrian model by evolutionary
  adjustment to video tracking Data}.
\newblock {\em Advances in Complex Systems}, 10(2):271--288, 2007.

\bibitem{Parisi2009}
Daniel~R. Parisi, Marcelano Gilman, and Herman Moldovan.
\newblock {A modification of the social force model can reproduce experimental
  data of pedestrian flows in normal conditions}.
\newblock {\em Physica A}, 388(17):3600--3608, 2009.

\bibitem{Moussaid2009}
M.~Moussa\"id, D.~Helbing, S.~Garnier, A.~Johansson, M.~Combe, and
  G.~Theraulaz.
\newblock {Experimental study of the behavioural mechanisms underlying
  self-organization in human crowds}.
\newblock {\em Proc. R. Soc. B.}, 276(1668):2755--2762, 2009.

\bibitem{Helbing2003}
D.~Helbing, M.~Isobe, T.~Nagatani, and K.~Takimoto.
\newblock {Lattice gas simulation of experimentally studied evacuation
  dynamics}.
\newblock {\em Phys. Rev. E}, 67:067101, 2003.

\bibitem{Bando1995}
M.~Bando, K.~Hasebe, A.~Nakayama, A.~Shibata, and Y.~Sugiyama.
\newblock {Dynamical model of traffic congestion and numerical simulation}.
\newblock {\em Phys. Rev. E}, 51(2):1035--1042, feb 1995.

\bibitem{Nakayama2005}
Akihiro Nakayama, Katsuya Hasebe, and Yuki Sugiyama.
\newblock Instability of pedestrian flow and phase structure in a
  two-dimensional optimal velocity model.
\newblock {\em Phys. Rev. E}, 71:036121, 2005.

\bibitem{Hirai1977}
K.~Hirai and K.~Tarui.
\newblock A simulation of the behavior of a crowd in panic.
\newblock {\em Systems and Control}, 21(6):409--411, 1977.

\bibitem{Karamouzas2009}
Ioannis Karamouzas, Peter Heil, Pascal van Beek, and Mark~H. Overmars.
\newblock A predictive collision avoidance model for pedestrian simulation.
\newblock {\em Lect. Notes Comp. Sc.}, 5884:41--52, 2009.
\end{thebibliography}

\end{document}